\def\VEL{\:{\rm km\:s^{-1}}}

\def\LA{Lyman\thinspace$\alpha$}

\def\HEiiL{\ion{He}{2} $\lambda1640$}

\def\NvL{\ion{N}{5} $\lambda\lambda1239,1243$}

\def\CiiLxiii{\ion{C}{2} $\lambda1333$}

\def\CiiiLxi{\ion{C}{3} $\lambda1176$}
\def\CivL{\ion{C}{4} $\lambda\lambda1548,1551$}

\def\SIiiiL{\ion{Si}{3} $\lambda\lambda1295-1303$}

\def\SIivL{\ion{Si}{4} $\lambda\lambda1394,1403$}

\newcommand{\MSOL}{\mbox{$\:M_{\sun}$}}

\newcommand{\EXPU}[3]{\mbox{\rm $#1 \times 10^{#2} \rm\:#3$}}  
\newcommand{\POW}[2]{\mbox{$\rm10^{#1}\rm\:#2$}}

\newcommand{\FUSE}{{\it FUSE}}
\newcommand{\HST}{{\it HST}}

\documentclass[iop]{emulateapj}
\slugcomment{Accepted for publication in The Astrophysical Journal}
\usepackage{amsmath,graphicx,float,ifthen,hyperref,times}

\newboolean{OnlineEdition}
\setboolean{OnlineEdition}{true} 

\graphicspath{{./figures/}}

\begin{document}

\shorttitle{Winds of RW Tri and UX UMa}
\shortauthors{Noebauer et al.}

\title{The Geometry and Ionization Structure of the Wind in the Eclipsing Nova-like Variables RW Tri and UX UMa}

\author{Ulrich M. Noebauer\altaffilmark{1, 2}, 
Knox S. Long \altaffilmark{1}, 
Stuart A. Sim \altaffilmark{2}, 
and Christian Knigge\altaffilmark{3}}

\altaffiltext{1}{Space Telescope Science Institute, 3700 San Martin Drive, Baltimore, MD 21218, USA}
\altaffiltext{2}{Max Planck Institute for Astrophysics, Karl-Schwarzschild-Str. 1, 85748 Garching, Germany}
\altaffiltext{3}{Department of Physics, University of Southampton, Southampton S017 1BJ UK}

\begin{abstract}

The UV spectra of nova-like variables are dominated by emission from
the accretion disk, modified by scattering in a wind emanating from
the disk.  Here, we model the spectra of RW Tri and UX UMa,  the only
two eclipsing nova-like which have been observed with the {\it Hubble
Space Telescope} in the far-ultraviolet, in an attempt to constrain the
geometry and the ionization structure of their winds.
Using our Monte
Carlo radiative transfer code, we computed spectra for
simply parameterized axisymmetric biconical outflow models and were
able to find plausible models for both systems.
These
reproduce the primary UV resonance lines -- \ion{N}{5}, \ion{Si}{4},
and \ion{C}{4} -- in the observed spectra in and out of eclipse.  The
distribution of these ions in the wind models is similar in both cases as
is the extent of the primary scattering regions in which these lines
are formed. The inferred mass-loss rates are 6\% -- 8\% of the mass
accretion rates for the systems. We discuss the implication of our
point models for our understanding of accretion disk winds in
cataclysmic variables.

\end{abstract}


\section{Introduction \label{sec:intro}}

Mass outflows are associated with accretion via a disk
onto compact objects  in a wide range of astrophysical systems,
ranging from proto-stellar objects to close binaries and active
galactic nuclei. The presence of mass outflow in all these systems,
which cover a wide range of scales and system parameters, suggests an
important connection between the processes of accretion and the
outflow. Cataclysmic variables (CVs) are the closest
astrophysical examples of such systems, and hence are an important
test case for our understanding of outflows from disks.  CVs are close
binary systems consisting of a white dwarf (WD) and a low-mass
late-type secondary star. The secondary overflows its Roche lobe,
leading to mass transfer to the WD. Due to angular momentum
conservation, the transferred material builds up an accretion disk
around the WD (unless the magnetic field of the WD is large enough to
disrupt the flow).  Disk emission usually dominates other sources of
light at visible wavelengths.  Many CVs undergo outbursts of 3 -- 5
mag lasting from days to weeks on timescales of weeks to years.  The
outbursts are due to a thermal instability in the disk which
transforms the disk from a relatively cold, largely neutral gas to
a hot, ionized plasma, increasing the accretion rate in the inner disk
from about  \EXPU{1}{-11}{\MSOL~yr^{-1}} to
$10^{-9}-10^{-8} \MSOL$ \mbox{yr$^{-1}$} \citep{Hoshi1979,
Mineshige1983}.  The frequency of outbursts depends on the longer term
mass-transfer rate from the secondary, which for systems that undergo
outbursts must be between the quiescent and outburst
rate \citep{Cannizzo1993}. 

Far-ultraviolet (FUV) spectra of CVs obtained with the {\it International Ultraviolet Explorer} ($IUE$) were the first to show that CVs
drive winds when the disk is in the high mass-transfer
state \citep{Krautter1981, Greenstein1982, Cordova1982}. Winds are not
observed when CVs are in the quiescent, low mass-transfer state. The
signature of the wind is imprinted on the spectra in the form of P
Cygni-like and/or blueshifted absorption profiles of resonance lines,
such as \ion{N}{5}, \ion{Si}{4}, and \ion{C}{4}. The blue edges of the absorption profiles
imply velocities exceeding 3000 $\VEL$ in some systems \citep[see, for
a recent example,][]{Hartley2002}.  The $IUE$ observations were used
to show that the  emission features are generally weak or absent in
systems viewed at low inclination and progressively more prominent at
higher inclination \citep{LaDous1991}.  Time-resolved  $IUE$
observations also demonstrated that line shapes in eclipsing and
nearly eclipsing systems vary systematically through secondary
conjunction, showing not only that the emission lines had to arise
from a region at least comparable in size to that of the secondary
star but also that the outflowing material was likely
rotating \citep{Cordova1985, Drew1988}.  While initial modeling of CV
winds favored spherical geometry for simplicity \citep{Drew1985,
Mauche1987}, the observed characteristics of the resonance lines in
CVs have over time led to a consensus picture of a biconical mass
outflow originating from the inner disk.   

\cite{Shlosman1993}, hereafter SV93, were the first to develop a
radiative transfer code based on the Sobolev approximation that was
capable of 
determining the ionization structure of a biconical outflow emerging
from the disk and of modeling \ion{C}{4} line profiles. They then used this
to attempt to constrain the wind geometry of non-magnetic
CVs. \cite{Vitello1993} found that they could reproduce the observed
profiles of \ion{C}{4} in three systems --- RW Sex, RW Tri, and V Sge --- with
plausible geometries and mass-loss rates;  they also found that they
could produce similar \ion{C}{4} profiles with quite different accretion
rates by changing other parameters in their kinematic description of
the wind.  Subsequently, \cite{Knigge1995}, hereafter KWD95, developed
another Monte Carlo radiative transfer code which solved the radiative
transfer exactly using a slightly different kinematic prescription for
a biconical wind. Their code did not calculate the ionization balance
in the outflow and instead assumed constant ionization fractions
throughout. As had \cite{Shlosman1993}, they were able to
reproduce qualitatively many of the basic characteristics of \ion{C}{4} line
profiles in disk-dominated CVs.  \cite{Knigge1997} applied their code
to high-resolution time-resolved {\it Hubble Space Telescope} (\HST) spectra of \ion{C}{4} of the eclipsing
nova-like variable UX UMa, concluding that one not only needed a fast
wind with a wide opening angle but a relatively dense, slow-moving
transition region between the photosphere of the disk and the wind.   

Despite the clear observational indications for the presence of mass
outflow in CVs, much is still unknown about the physics of these
winds, in particular the driving mechanisms. By analogy with O-star
winds, most have assumed that the winds of CVs are
radiatively driven. Hydrodynamical simulations seem to broadly support
this picture \citep{Proga2002}, even though questions regarding
whether CV disks are luminous enough to drive the required mass
outflow remain \citep{Drew2000}.  Alternatives, magnetically driven or
magnetically assisted flows from the disk or boundary layer region
exist, based on suggestions that can be traced to, among
others, \cite{Blandford1982} and \cite{Pringle1989}, and are widely
discussed for other disk systems, ranging from young stellar objects 
to active galactic nuclei \citep[see,
e. g.,][]{Koenigl2010, Campbell2010}. An important obstacle to a
confident identification of the driving mechanism is that even basic
wind parameters, such as the mass-loss rate, the
acceleration, collimation, and ionization structure, remain poorly constrained. Thus 
obtaining reliable information on the wind structure, and especially
about the ionization in the wind, should lead to a much better
understanding of the physics underlying CV winds. In the long term,
this might even allow us to gain insight into other types of accreting
systems, as mass outflows seem to be a general feature of such systems. 

Nova-like variables are disk-dominated CVs, which, unlike most other
CVs, are nearly always observed in the ``outburst" state. This is presumably
because they have a higher mean mass-transfer rate than other types of
CV (e.g., dwarf novae; \citealt{Schreiber2002}).  Nova-like variables have orbital periods just above 
the period gap in non-magnetic CVs, where magnetic braking is
thought to produce the largest mass-transfer rates in the long-term
evolution of CVs \citep{Howell2001}.  Because they stay in the high
mass-transfer state for long periods of time, they may be the closest
examples to steady-state accretion disks among the CVs.  RW Tri and UX
UMa are the two brightest eclipsing examples of this class.  Eclipsing
systems are of particular interest in the study of CVs, and, in
particular, the winds of CVs, since time-resolved observations of the
eclipse provide spatial information about the disk and wind properties
that is unavailable for systems viewed at lower inclinations.   

Here, we describe our efforts to model the \HST\ spectra of RW Tri and
UX UMa in an attempt to place constraints on the wind geometry and the
ionization structure. These are the  only eclipsing nova-like CVs
for which medium spectral resolution, time-resolved UV spectra
exist. The observations of RW Tri have been briefly discussed
by \cite{Mason1997}; the observations of UX UMa have been discussed
more extensively
by \cite{Mason1995}, \cite{Mason1997}, \cite{Knigge1997}, \cite{Baptista1998}, \cite{Froning2003},
and  \cite{Linnell2008}.   With the partial exception
of \cite{Knigge1997}, the earlier studies were primarily focused on
modeling the accretion disk and WD in UX UMa.   However, no one has
attempted to constrain the properties of the wind in these systems by
modeling all of the prominent wind lines simultaneously.  The
remainder of this paper is organized as follows. In Section \ref{sec:obs},
we describe the time-resolved observations of RW Tri and UX UMa with
the Goddard High Resolution Spectrograph (GHRS)  and the selection of the spectra we attempt to model.
In Section \ref{sec:code}, we describe briefly the radiative transfer code
developed by \cite{Long2002} to produce synthetic spectra from
parameterized models of biconical outflows, which we use to generate
synthetic spectra. Then, in Section \ref{sec:approach}, we describe the strategy we used to identify models for the systems, and in Section \ref{sec:results},
we discuss the degree to which we were able to reproduce the observe
spectra and the implications of the results on  the wind geometry of
RW Tri and UX UMa.  Finally, in Section \ref{sec:conclusions}, we summarize
what we have learned through our attempt to model the pre-eclipse and
eclipse spectra of RW Tri and UX UMa.

\section{Observations}
\label{sec:obs}

UV spectra of the nova-like RW Tri and UX UMa were obtained with the
GHRS instrument on \HST\  in late 1996 and early 1997 (as part of
programs 6494 and 6024) and are described by \cite{Mason1997}. Both
systems were observed in order to explore time variations in the
spectra as a function of the orbital phase.  All of the observations
discussed here were performed with the G140L grating, providing a
wavelength resolution of 0.57 \AA, in the rapid read-out mode, in which
spectra were recorded at 5 s intervals. Each observation covers one of 
two overlapping wavelength intervals, each with a bandwidth of about
287 \AA\, so that spectra from separate observations must be combined
to yield the full spectral range from 1150 \AA\  to
1660 \AA. The observations of  UX UMa and RW Tri each comprised about
nine \HST\ orbits. An observation log is presented in
Table \ref{tab:Obs_overview}. To carry out this study, we
retrieved and used the final reprocessed data from the \HST\ archive. 

\begin{table}[t]
  \centering
  \caption{Observation Log for \HST\ Spectra of  RW Tri and UX UMa.}
  \begin{tabular}{@{}l l c l@{}}
    \hline\hline
    \multicolumn{1}{c}{ObsID} & \multicolumn{1}{c}{Obs. Start (MJD)} & \multicolumn{1}{c}{Spectral range (\AA)} & \multicolumn{1}{c}{Phase Coverage} \\
    \hline
    \multicolumn{4}{c}{UX UMa}\\
    \hline
    z3fy0304t   &50398.2014& 1148-1435 & ~0.654 - 0.701\\
    z3fy0306t   &50398.2132& 1376-1663 & ~0.714 - 0.762\\
    z3fya102t * &50395.1085& 1148-1435 & -0.073 - 0.033\\
    z3fya302t   &50398.2554& 1148-1435 & -0.071 - 0.093\\
    z3fyb102t   &50395.1751& 1376-1663 & ~0.027 - 0.377\\
    z3fyb104t   &50395.1998& 1148-1435 & ~0.391 - 0.431\\
    z3fyb106t   &50395.2422& 1148-1435 & ~0.607 - 0.647\\
    z3fyb108t   &50395.2526& 1376-1663 & ~0.660 - 0.771\\
    z3fyb10at * &50395.3091& 1376-1663 & ~0.053 - 0.111\\
    \hline
    \multicolumn{4}{c}{RW Tri}\\
    \hline
    z3kxa102t   &50471.3006& 1148-1435& ~0.395 - 0.433\\
    z3kxa104t   &50471.3446& 1148-1435& ~0.585 - 0.718\\
    z3kxa106t * &50471.4116& 1148-1435& -0.126 - 0.007\\
    z3kxa202t   &50470.6978& 1148-1435& ~0.796 - 0.893\\
    z3kxa204t   &50470.7418& 1148-1435& -0.015 - 0.118\\
    z3kxa302t   &50474.3154& 1148-1435& ~0.395 - 0.433\\
    z3kxa304t   &50474.3593& 1376-1663& ~0.585 - 0.718\\
    z3kxa306t * &50474.4263& 1376-1663& -0.126 - 0.007\\
    z3kxa402t   &50473.7126& 1376-1663& ~0.796 - 0.833\\
    z3kxa404t   &50473.7566& 1376-1663& -0.015 - 0.118\\
    z3kxb102t   &50471.4786& 1376-1663& ~0.162 - 0.296\\
    z3kxb302t   &50474.4933& 1376-1663& ~0.163 - 0.296\\
    \hline

  \end{tabular}

  \label{tab:Obs_overview}
  \begin{flushleft}
    {\bf Note.} Average spectra are constructed from observations labeled with an asterisk (*).
  \end{flushleft}
\end{table}

\begin{figure}[t]
  \centering
  \ifthenelse{\boolean{OnlineEdition}}{
    \includegraphics[width=0.48\textwidth]{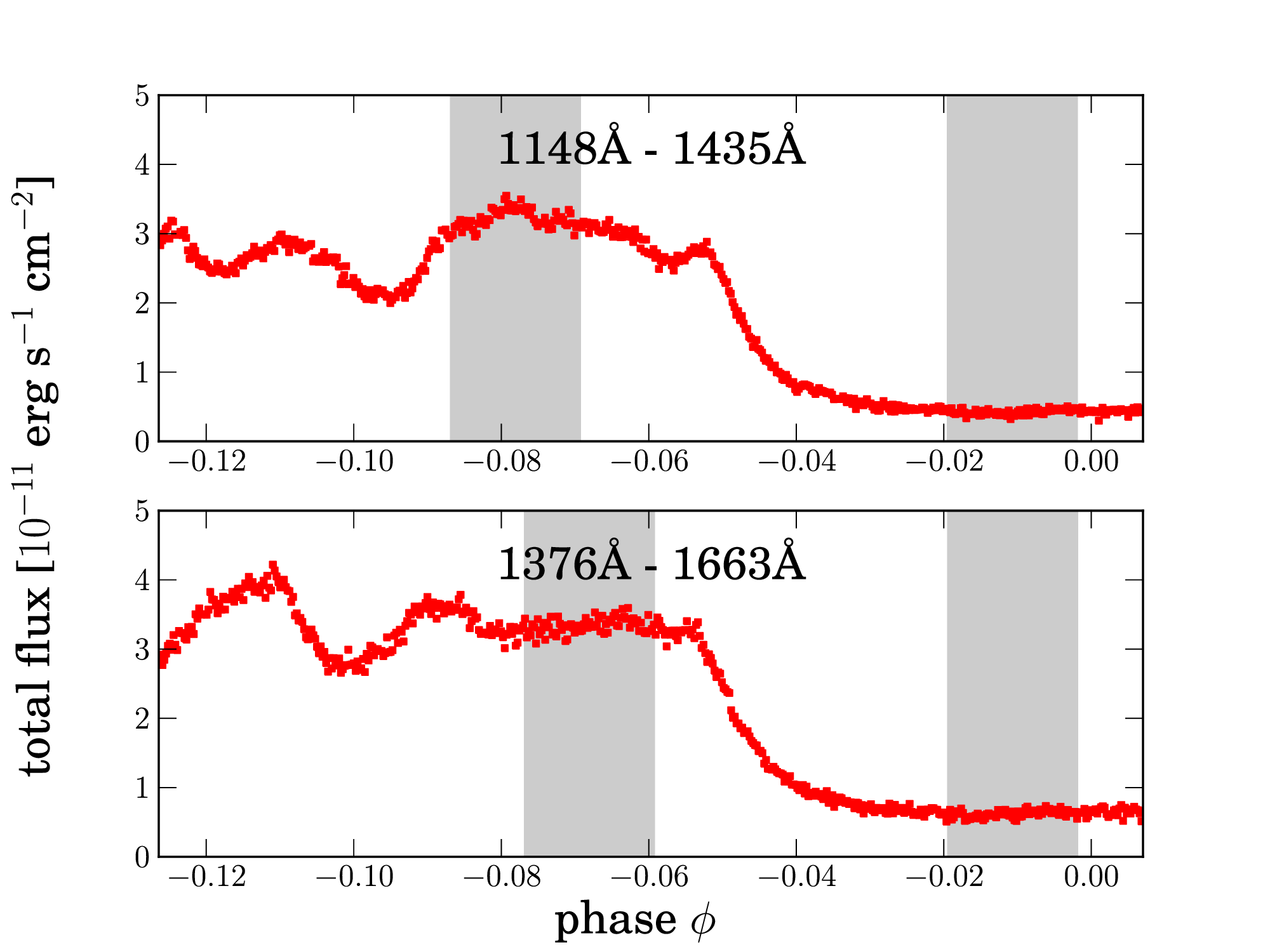}
  }
  {
    \includegraphics[width=0.48\textwidth]{rwtri_lightcurve_1_bw.eps}
  }
  \caption{Light curves of the two RW Tri observations that were used to construct average pre-eclipse and mid-eclipse spectra. The phase of each measurement was derived with the revised ephemeris of RW Tri of \cite{Groot2004}. The measurements in the upper panel all cover the shorter wavelength band from 1148 \AA\ to 1425 \AA, the observations in the lower panel were taken in the band from 1376 \AA\ to 1663 \AA. The phase intervals that were used for the averaging process are shaded in gray.  
}
  \label{fig:rwtri_obs_lc}
\end{figure}
\begin{figure}[t]
  \centering
  \ifthenelse{\boolean{OnlineEdition}}{
    \includegraphics[width=0.48\textwidth]{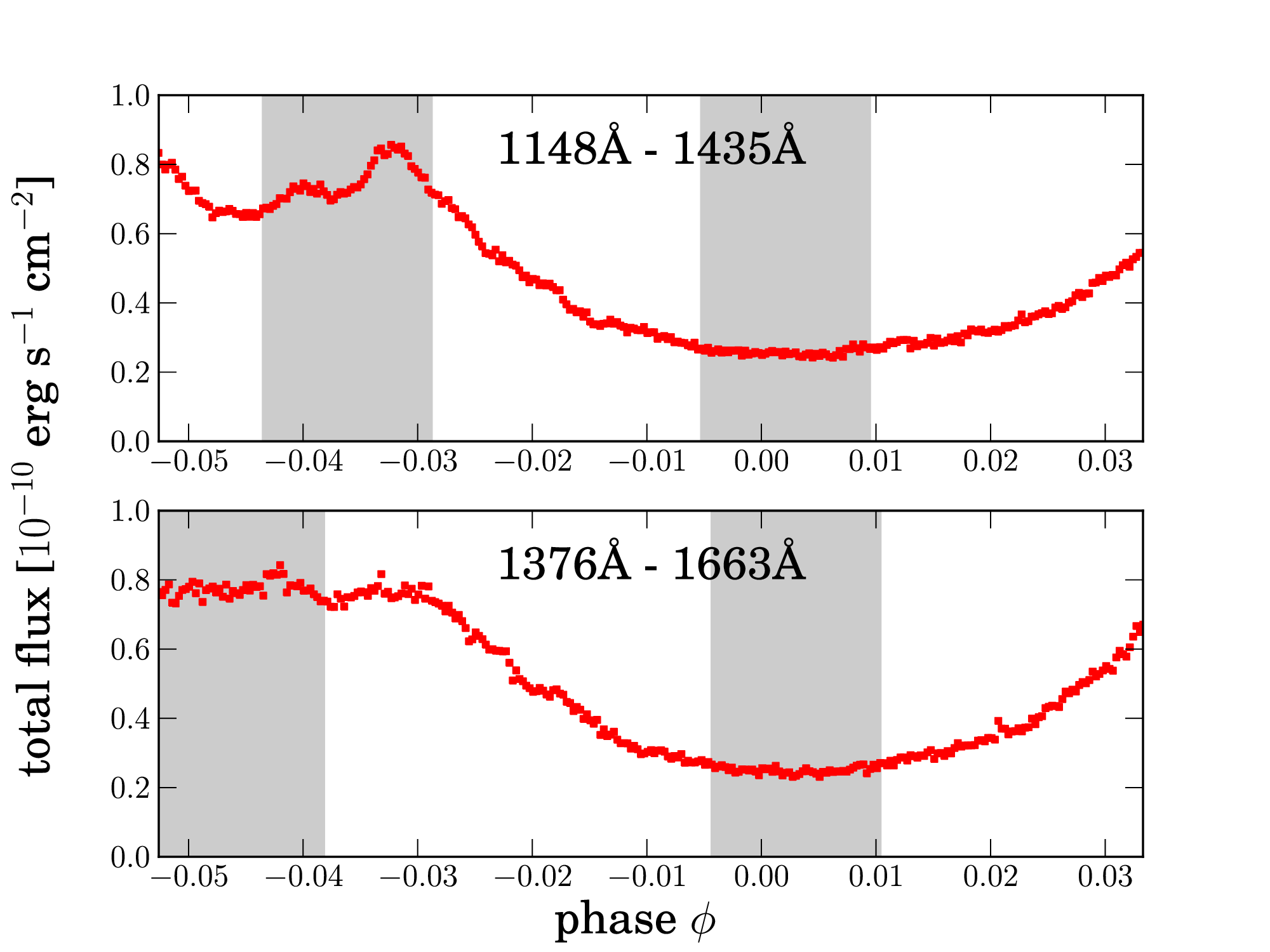}
  }
  {
    \includegraphics[width=0.48\textwidth]{uxuma_lightcurve_1_bw.eps}
  }
  \caption{Light curves of UX UMa. The phase of each measurement was determined with the ephemeris of UX UMa derived by \cite{Baptista1995}. The upper panel shows the observation in the short wavelength band. In the lower panel, the observation covering the long wavelength band is displayed. The average spectra were constructed from the shaded regions. 
   }
  \label{fig:uxuma_obs_lc}
\end{figure}
\begin{figure}[t]
  \centering
  \ifthenelse{\boolean{OnlineEdition}}{
    \includegraphics[width=0.48\textwidth]{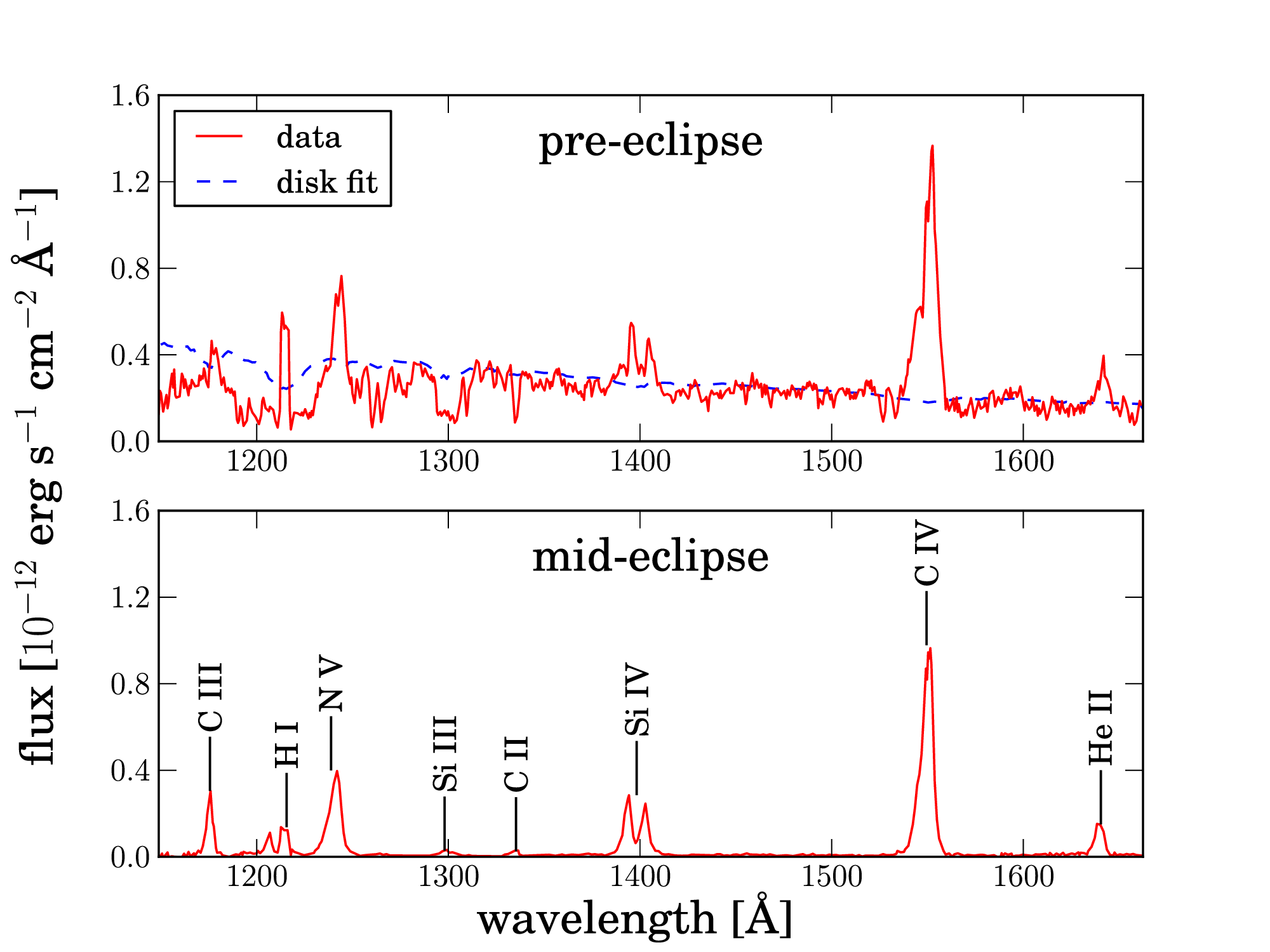}
  }
  {
    \includegraphics[width=0.48\textwidth]{rwtri_data_1_bw.eps}
  }
  \caption{Average RW Tri \HST\ pre-eclipse and mid-eclipse spectra obtained by averaging over the regions highlighted in Figure \ref{fig:rwtri_obs_lc}. In the pre-eclipse panel, the result of fitting the continuum to an accretion disk radiating as an ensemble of stellar atmospheres is also shown. The main emission lines are shown in the lower panel.
  }
  \label{fig:rwtri_obs_spec}
\end{figure}
\begin{figure}[t]
  \centering
  \ifthenelse{\boolean{OnlineEdition}}{
    \includegraphics[width=0.48\textwidth]{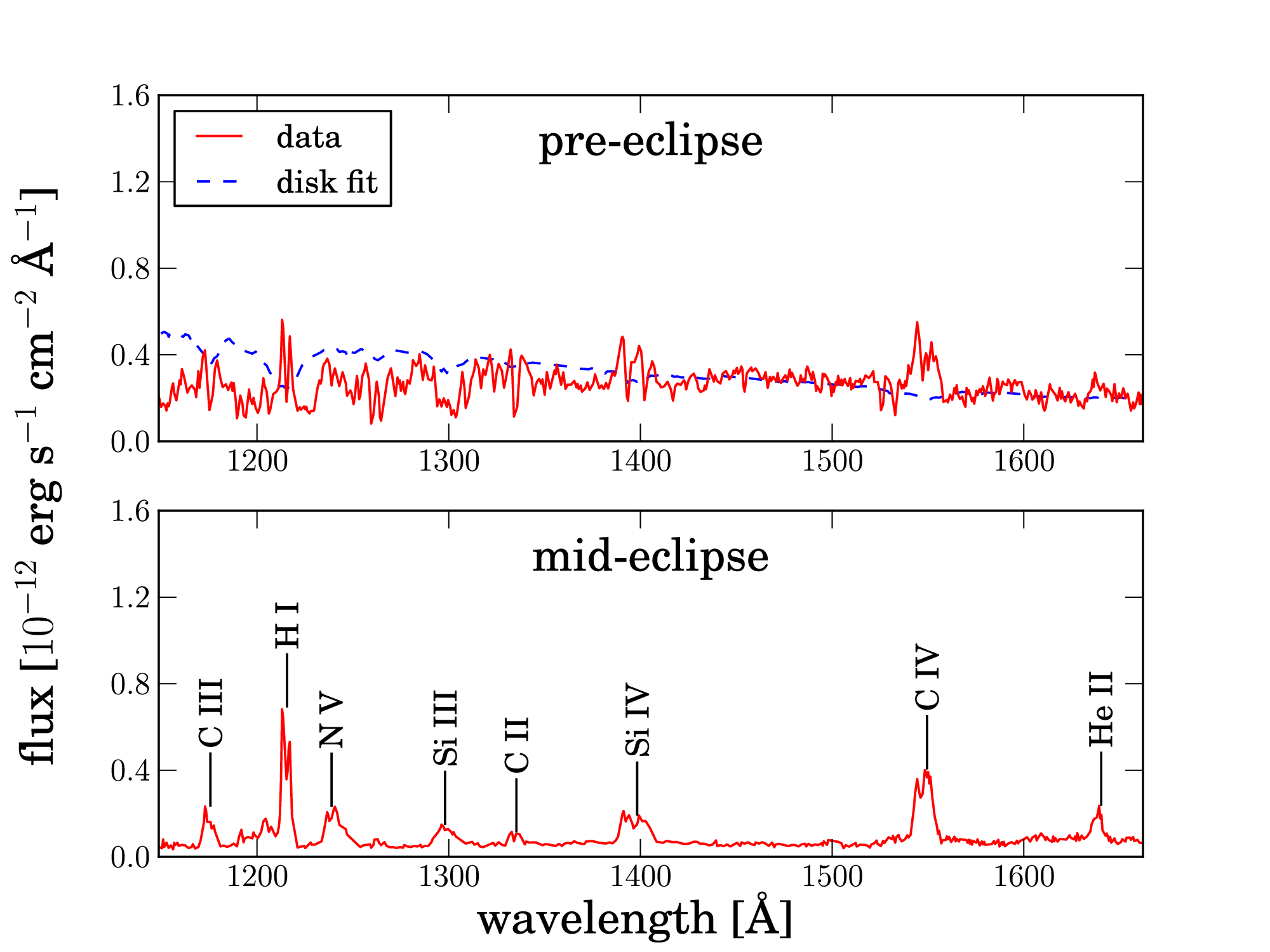}
  }
  {
    \includegraphics[width=0.48\textwidth]{uxuma_data_1_bw.eps}
  }
  \caption{Average UX UMa \HST\ pre-eclipse and mid-eclipse spectra. Again emission lines are identified in the lower panel and the disk continuum fit is shown in the upper panel.}
  \label{fig:uxuma_obs_spec}
\end{figure}

To facilitate comparison of the synthetic spectra to the actual spectra of RW Tri
and UX UMa, we averaged the observed spectra over specific phase
intervals. There were certain difficulties associated with this. At any time 
observations only covered parts of the spectral range of interest and the
phase coverage was also not ideal for either UX UMa or
RW Tri, as indicated in Table 1. For RW Tri, two observations sampled
pre-eclipse and mid-eclipse orbital phase. The other observations
covered only parts of the mid-eclipse and the post-eclipse state. In
both cases, one observation existed for each wavelength band. For UX
UMa, there was only one observation covering the eclipse in the 1375 -- 1665 \AA\ 
band, but two covering the shorter wavelength band from 1148 \AA\ to 1435 \AA.
To construct representative spectra for the mid-eclipse and the pre-eclipse phase,
both RW Tri observations containing information of these two states
were taken. For UX UMa, the two observations in the 1148 --
1435 \AA\ band had different overall pre-eclipse flux levels but showed
no striking differences in spectral shape. Therefore, only the observation
which was closest to the overall flux level of the only measurement
set in the longer wavelength band was used. The spectra used in our
study were obtained by averaging over representative phase
intervals. Choosing the width and the position of these intervals was,
at least in the pre-eclipse case, somewhat arbitrary, but tests showed
that the overall flux level and the spectral features in the average
spectra were not very sensitive to these choices. In the end, we used
the highlighted regions in Figures \ref{fig:rwtri_obs_lc}
and \ref{fig:uxuma_obs_lc} for the averaging process. The resulting
spectra were finally merged to cover the entire wavelength band from
1148 \AA\ to 1665 \AA\ by averaging over the overlapping
region. Figures \ref{fig:rwtri_obs_spec}  and \ref{fig:uxuma_obs_spec}
show the resulting pre-eclipse and eclipse spectra for UX UMa and RW
Tri, respectively. According to \cite{Rutten1992}, RW Tri is thought
to lie along a line of sight with {\it E(B--V)} of 0.1, and so to facilitate
comparisons with models the spectra for RW Tri were corrected for
reddening. Reddening along the line of sight to UX UMa is believed
small ({\it E(B--V)}$<$0.04, \citealt{Verbunt1987}), so no correction was made
for UX UMa. 

The pre-eclipse spectra of both RW Tri and UX UMa show rather flat
continua with a number of superimposed resonance lines. During eclipse the
continuum is suppressed and the spectrum is dominated by the emission
lines that, because the disk is eclipsed, must arise from an extended
region.

The three major resonance lines present in both systems are
the doublets \NvL, \SIivL,  and \CivL. Apart from these there are
additional, weaker lines of low-ionization state metals
such as \CiiLxiii, \CiiiLxi, and \SIiiiL\ and lines of both hydrogen
(\ion{H}{1} \LA\ $\lambda1215$) and helium (\HEiiL) are visible. In the
pre-eclipse state of both systems, the low ionization state lines,
especially those of \CiiLxiii\ and \SIiiiL\ are seen as deep absorption dips, suggesting that the main
concentration of these ions lies in the line of sight. The UX UMa
pre-eclipse spectra generally show many absorption features, including in
the main resonance lines: for example, two prominent narrow dips are present in
the \ion{Si}{4} line and one in the \ion{C}{4} line.  

\section{Radiative transfer code and wind models}
\label{sec:code}

In order to place limits on the geometry of the winds in RW Tri and UX
UMa using the spectra, we need to model the effect  of the wind on the spectra. Our 
working hypothesis is that the winds arise in a
biconical outflow.  To model the spectra we parameterize the outflow using one of 
two kinematic descriptions for the wind and we use a Monte
Carlo radiative transfer code  ({\sc python}, developed
by \citealt{Long2002}, with improvements described by \citealt{Sim2005}) to
calculate the ionization structure of the wind and to create synthetic
spectra. The chemical composition of the wind material is fixed to
the solar abundances determined by \cite{Anders1989}.
Here, we give a short summary of the important features of
this program and the two classes of parameterized wind models which we use
in this study (for full details, see \citealt{Long2002}).

\subsection{Monte Carlo approach}

{\sc python} operates by performing a sequence of Monte Carlo
simulations in which the quanta are packets of radiative
energy. These packets are launched in accordance with the specified
radiation sources (see below) and propagate through a parameterized
model for the CV disk wind. As the packets propagate, they can
interact with the wind material via electron scattering, resonance
line scattering, bound--free absorption, or free--free absorption. Their
trajectories are finally terminated when they either escape, lose all
their energy to absorption in the wind, or are lost from the simulation
by striking the WD, disk surface, or companion star. 

Since the ionization and heating of the wind are controlled by the
radiation, the ionization and thermal state must be computed
self-consistently with the local radiation field properties. This is
achieved via an iteration sequence. In each iteration, the wind
properties are held fixed while a Monte Carlo simulation of the
radiation field is performed. During the Monte Carlo simulation,
estimators for the local radiation field properties (radiation
temperature $T_{\rm R}$ and dilution factor $W$) are recorded for each
grid cell in the wind and used to make an improved estimate of the
wind ionization state and kinetic temperature $T_{\rm e}$ for use in the
next iteration.  

Once the wind properties reach convergence, a final sequence of
additional Monte Carlo simulations for the converged wind properties
are used to obtain the observed spectrum for the desired inclination
angle. In these final simulations, a modified version of the technique
described by \cite{Woods1991} is used: in every interaction with the
wind medium the packet contributes to the final spectrum according to
the probability of emission into the observer's line of sight after
the interaction. 

\subsubsection{Radiation sources}

The calculations presented in this work include radiation by the WD
and its accretion disk. The disk is the primary source of radiation in 
non-magnetic CVs in the high mass-transfer state. It is treated as a 
geometrically thin but optically thick disk which follows the standard 
temperature distribution (e.g., \citealt{Wade1984}) 

\begin{equation}
  \label{eq:effective_temp}
 T_{\rm eff}=\left( \frac{3 G \dot M_{\rm acc} M_{\rm WD}}{8 \pi \sigma R_{\rm WD}^3}\right)^{1/4} \left(\frac{R_{\rm WD}}{R}\right)^{3/4} \left(1-\sqrt{\frac{R_{\rm WD}}{R}}\right)^{1/4}.
\end{equation}

The disk radiation is composed from an ensemble of 300 concentric annuli
each of which radiates a spectrum at the local value of $T_{\rm eff}$. The 
inner and outer boundaries of each annulus are determined by the requirement that
all annuli should emit an equal fraction of the total disk luminosity.
For this study, we model the spectrum from each disk annulus as a black body 
during the iteration processes used to obtain the ionization state. In the 
spectrum calculation, the disk emission is described by synthetic stellar 
spectra created with TLUSTY/SYNSPEC \citep{Hubeny1995}. When synthetic spectra
are used, the calculation of the effective gravity of the disk follows the 
prescription given by \cite{Herter1979}.
To describe the WD radiation, a single black body/Hubeny model is
used. The secondary star is treated as a dark absorber. 

The {\sc python} code can also
account for radiation from a boundary layer parameterized by a
luminosity and temperature as described by \cite{Long2002}. We found,
however, that if we included a hot luminous boundary layer (such as
introduced by \citealt{Knigge1997}) the wind was typically too highly
ionized to obtain a good match with observed spectra. We therefore
opted to exclude any such boundary layer from the calculations. This
may imply that either the WD has significant rotation or that the
boundary layer geometry is such that much of the wind is not exposed
to its radiation.

The energy emitted by the radiation sources is discretized into energy
packets, each with its own direction of propagation, frequency, and
weight. The packets are initialized with weights corresponding to the
fraction of the physical luminosity carried by the package and
frequency sampled randomly from the luminosity spectra of the WD and
accretion disk annuli. Initial directions of propagation are assigned
assuming a linear limb-darkening law and initial positions are
located on either the WD or disk surface, as appropriate.  
In addition to the primary radiation sources, the code also accounts for free--free, free--bound, and bound--bound emission by the wind, based on its local ionization state and kinetic temperature.

\subsubsection{Wind conditions}

A key component of modeling the wind spectrum is a realistic treatment of the wind ionization state. In contrast to some previous studies in which uniform ionization conditions throughout the wind were assumed (e.g., \citealt{Knigge1995}), {\sc python} computes the ionization state as a function of position using the modified nebular approximation described by \cite{Mazzali1993}

\begin{equation}
  \label{eq:ionization_balance}
  \frac{n_{j+1}n_{\rm e}}{n_j} = W [\xi + W(1-\xi)]\left( \frac{T_{\rm e}}{T_{\rm R}}\right)^{1/2} \left(\frac{n_{j+1}n_{\rm e}}{n_j}\right)^{*}~.
\end{equation}
Here, $W$ is a dilution factor, $\xi$ is the fraction of recombinations
that go directly to the ground state, and $T_{\rm e}$ and $T_{\rm R}$ are the
electron and radiation temperatures, respectively. The asterisk
denotes quantities to be evaluated in local thermodynamic equilibrium at $T_{\rm R}$. 

In contrast to \cite{Mazzali1993}, however, we do not assume a simple
relationship between $T_{\rm R }$ and $T_{\rm e}$ but instead calculate $T_{\rm e}$
independently from the assumption of thermal equilibrium (i.e., the
heating and cooling rates are in equilibrium at every point in the
wind). As noted above, $T_{\rm R}$, $W$, and $T_{\rm e}$ are not known a
priori, but are determined by an iterative sequence of Monte Carlo
experiments. In practice, $T_{\rm R}$ is the most important parameter for
the ionization balance (given by Equation (\ref{eq:ionization_balance})) and is
primarily determined by the temperature distribution in the
disk. This in turn depends on the adopted system parameters (WD mass
and radius, and the mass accretion rate $\dot M$).  

\subsection{Wind models}

For our simulations, we adopt two classes of simply parameterized wind
models, those developed by SV93 and KWD95. We carried out our studies using both wind models so that we could investigate any sensitivity of our results to the parameterization. As demonstrated below, however, we find that our conclusions are relatively insensitive to this choice. Both of these models describe the wind as a stationary biconical outflow emerging from both sides of
the accretion disk. In each case, the velocity in the wind consists of
a rotational component $v_{\phi}$ and a poloidal component $v_{\rm q}$. The
rotational component is obtained assuming that the wind streamlines
rise from a disk in Keplerian rotation and that specific angular
momentum is conserved as the flow moves outward: 

\begin{equation}
  \label{eq:ang_momentum}
  v_{\phi}(r) r= v_{(\phi,0)}(R) R~,
\end{equation}

\begin{equation}
  \label{eq:kepler}
  v_{(\phi,0)}(R)=\sqrt{\frac{G M_{\rm WD}}{R}}~.
\end{equation}
Here, and in the following, $r$ stands for the radial position of a point in the wind, while $R$ denotes the radius at which the wind streamline passing through the point reaches the accretion disk.
The poloidal velocity field in the wind causes material to spiral on
streamline cones away from the disk. The detailed geometry of these
cones depends on which of the wind prescriptions is used. This,
together with the way of parameterizing the poloidal velocity and
calculating the wind density, is the main difference between the two
classes of wind models we consider. 

\subsubsection{KWD95 wind model}
\label{sec:kwd}

\begin{figure}[t]
  \centering
  \includegraphics[width=0.48\textwidth]{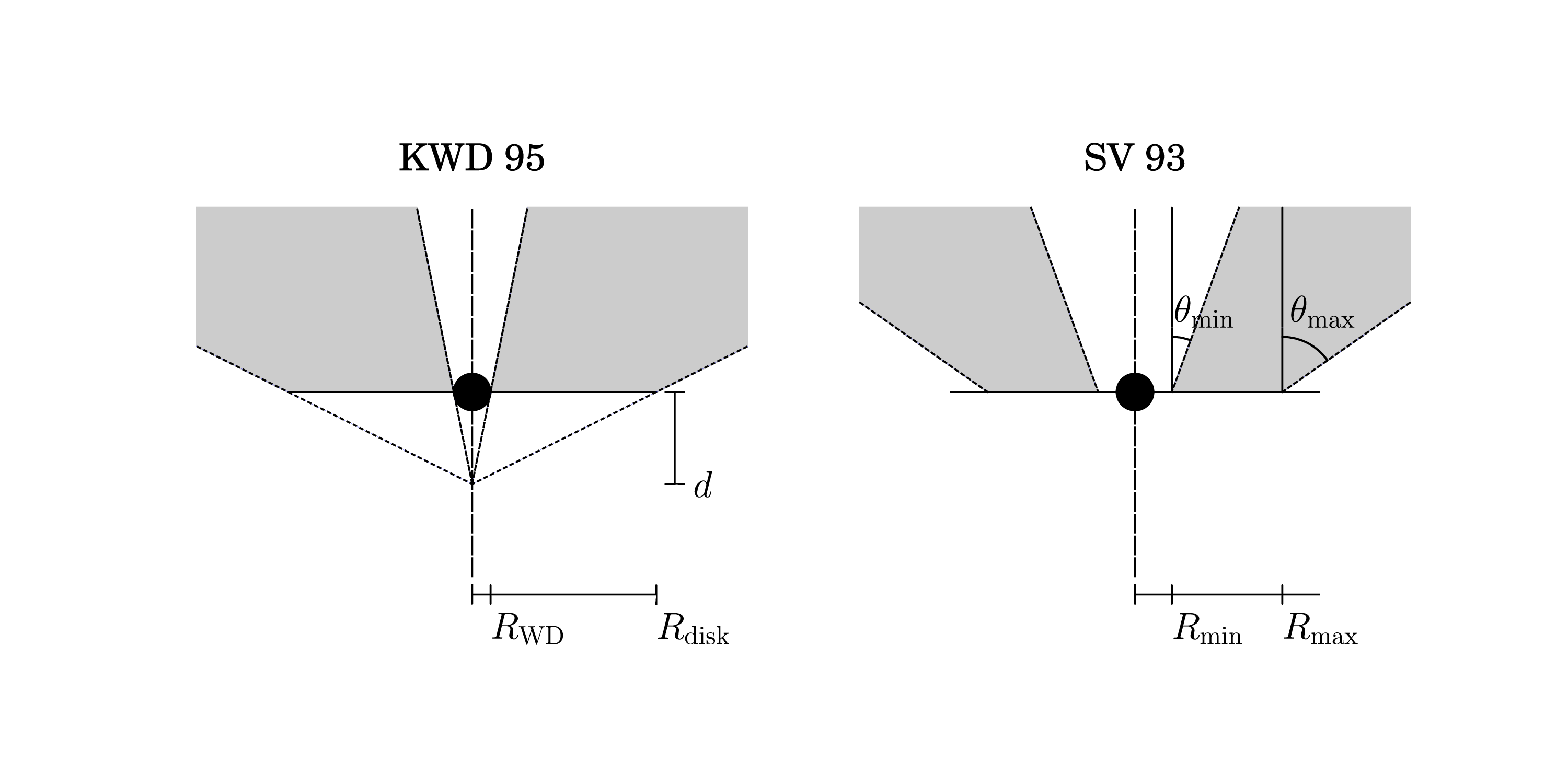}
  \caption{Left: wind geometry as parameterized by KWD95. The black circle denotes the WD and the horizontal line the accretion disk. Note that the wind shape is controlled by only one collimation parameter, $d$, which is usually given in units of $R_{\rm WD}$.
     Right: wind geometry as parameterized by SV93. In both panels, the mass outflow region is shaded in gray. For illustrative purposes only one half of the wind structure is shown (the wind is symmetric under reflection in the $xy$-plane).}
  \label{fig:geometry}
\end{figure}

KWD95 parameterize the wind as a biconical outflow from the entire
accretion disk and characterize its geometry by only one parameter
$d$. By definition, this parameter specifies the position of a point
on the rotational axis where extrapolations of all streamline cones
meet. Together with the disk extent $[R_{\rm WD},R_{\rm disk}]$, $d$ defines
the opening angles of the wind. Figure \ref{fig:geometry} illustrates
the wind geometry of the KWD95 treatment. 

The wind region itself is characterized by a density and a
velocity. The density in the wind is specified by the mass-loss rate
per area in the disk. This quantity is assumed to follow the
temperature distribution in the disk 

\begin{equation}
  \label{eq:kwd_mass_loss_rate}
  \dot m(R) = \dot M_{\rm wind} \frac{T_{\rm eff}^{4\alpha}(R)}{\int d A'  T_{\rm eff}^{4\alpha}(R')}~.
\end{equation}
The integral in the denominator ensures the normalization to the total
mass-loss rate $\dot M_{\rm wind}$  when integrated over the disk surface
from which the wind emerges. The exponent $\alpha$ allows the
mass-loss rate per unit area to be tied to the local disk effective
temperature ($\alpha=0.25$) or the luminosity ($\alpha=1.0$) in the
disk. The mass-loss rate per area can be used together with the
poloidal velocity $v_{(\rm q,0)}$ at the streamline base and the polar
angle of the streamline $\delta = \arctan \frac{R}{d}$ to calculate
the density at the base of the streamline

\begin{equation}
  \label{eq:kwd_density_base}
  \rho_0(R)=\frac{\dot m(R)}{v_{(\rm q,0)}(R) \cos \delta}~.
\end{equation}
As the wind rises from the disk, the density decreases as

\begin{equation}
  \label{eq:kwd_density_wind}
  \rho(r)=\rho_0(R)\left(\frac{d}{q \cos \delta}\right)^2 \frac{v_{(\rm q,0)}(R)}{v_{\rm q}(r)}~.
\end{equation}
Here $q$ denotes the distance to the point $d$ on the rotational
axis. The poloidal velocity component $v_{\rm q}(r)$ depends on three
parameters, the acceleration length $R_{\rm s}$, acceleration exponent
$\beta$, and terminal velocity $v_{\infty}$:

\begin{equation}
  \label{eq:kwd_velocity_wind}
  v_{\rm q}(l)=c_{\rm s}(R) + [v_{\infty} -  c_{\rm s}(R)]\left( 1 -  \frac{R_{\rm s}}{R_{\rm s}+l} \right)^{\beta}~.
\end{equation}
The terminal velocity is parameterized as a multiple ($f$) of the local escape speed

\begin{equation}
  \label{eq:kwd_velocity_terminal}
  v_{\infty}= f v_{\rm esc} = f \sqrt{\frac{2 G M_{\rm WD}}{R}}
\end{equation}
and the poloidal velocity is assumed to be equal to the local sound
speed at the base of the wind streamlines. 

\begin{equation}
  \label{eq:kwd_velocity_sound}
  c_{\rm s}(R)=10 \cdot \sqrt{\frac{T_{\rm eff}(R)}{10^4}}\mbox{km s$^{-1}$}~.
\end{equation}
The quantity $l$ in Equation (\ref{eq:kwd_velocity_wind}) represents the
distance of the point $r$ to the accretion disk, measured on the
streamline cone. 

In total there are six parameters, $d$, $\alpha$, $\beta$, $\dot
M_{\rm wind}$, $f$, and $R_{\rm s}$ which specify the CV wind in the KWD95
parameterization. Although the true geometry is controlled only by $d$,
the extent of the wind region in which the ion concentrations are
sufficiently high to lead to strong spectral lines is strongly
sensitive to the distribution of mass outflow from the disk, as determined
by $\dot M_{\rm{wind}}$ and $\alpha$. In particular, high values of the
exponent $\alpha$ will cause a very high density near the rotational
center of the wind and a very dilute outflow at the outer edges of the
wind while decreasing $\alpha$ moves more material to the outer
edges.  Thus, the effective collimation is determined by a combination
of $d$ and $\alpha$. 

\subsubsection{SV93 wind model}
\label{sec:sv}

SV93 presented a wind model that is more flexible than the KWD95
treatment, but involves more free parameters. In the SV93 description,
the wind does not necessarily emerge from all parts of the
disk. Instead, the wind region is specified by a minimum and a maximum
disk radius $R_{\rm min}$ and $R_{\rm max}$. Similarly, the collimation of the
wind is specified by a minimum and a maximum streamline angle
$\theta_{\rm min}$ and $\theta_{\rm max}$. Figure \ref{fig:geometry} illustrates
the wind geometry in the SV93 model. The distribution of streamline angles throughout the disk is determined by an additional parameter $\gamma$ defined via 

\begin{equation}
  \label{eq:sv_theta}
  \theta(R) = \theta_{\rm max} + (\theta_{\rm max} - \theta_{\rm min}) x(R)^{\gamma}~,
\end{equation}
where

\begin{equation}
  \label{eq:sv_streamline_pos}
  x(R)=\frac{R-R_{\rm min}}{R_{\rm max}-R_{\rm min}}.
\end{equation}
As in the KWD95 description, $R$ is the radius of the point at which
the streamline passing through point ${\bf r}$ crosses the disk
plane. 

The rotational velocity in the SV93 wind model is exactly the same as
in the KWD95 description (see Equations (\ref{eq:ang_momentum})
and (\ref{eq:kepler})). The poloidal velocity ($v_{\rm q}$) at a distance
$l$ measured on the streamline cone is specified in terms of three
parameters: an acceleration length scale $R_{\rm s}$, acceleration exponent
$\alpha$, and the terminal velocity $v_{\infty}$

\begin{equation}
  \label{eq:sv_velocity_wind}
  v_{\rm q}(l)=v_0 + (v_{\infty} - v_0) \left[ \frac{(l/R_{\rm s})^{\alpha}}{(l/R_{\rm s})^{\alpha}+1} \right]~.
\end{equation}
We follow SV93 in adopting $v_0 = 6$~km~s$^{-1}$. As in the KWD95 prescription,
the terminal velocity is specified in units of the local escape speed
(see Equation~(\ref{eq:kwd_velocity_terminal})). 

Again, the wind density is determined by the disk mass-loss rate per
unit area $\dot m$. In the SV93 prescription, this is not specified in
terms of the local disk temperature but is assumed to follow the
relation 

\begin{equation}
  \label{eq:sv_mass_loss_rate}
  \dot m (R) = \dot M_{\rm wind} \frac{R^{\lambda} \cos \theta(R)}{\int d A'  R'^{\lambda} \cos \theta(R')}~.
\end{equation}
The normalization ensures that calculating the total mass-loss rate is
given by $\dot M_{\rm wind}$ while $\lambda$ allows for variation in the
mass loading of wind streamlines. The density in the wind is then given by  

\begin{equation}
  \label{eq:sv_density_wind}
  \rho(r,z)=\frac{R}{r} \frac{ d R}{d r} \frac{\dot m (R)}{v_{\rm z}(r,z)}.
\end{equation}
The $R/r ~d R /  d r$ factor results from the increasing area between
streamlines the further the wind is away from the disk which, in the
special case of streamlines on a cone with angle $\theta$, simplifies
to  

\begin{equation}
  \label{eq:sv_density_par}
  \frac{ d r}{ d R} = 1 + \frac{l ( d \theta / d R)}{\cos \theta}.
\end{equation}
As in the KWD95 description, $l$ is the distance to the disk on the
streamline cone. 

In total, the SV93 prescription requires ten parameters to define the wind: 
$\gamma$, $R_{\rm min}$, $R_{\rm max}$, $\theta_{\rm min}$, $\theta_{\rm max}$,
$\lambda$, $R_{\rm s}$, $\alpha$, $f$, and $\dot M_{\rm wind}$.  The collimation
of the wind is primarily determined by $\theta_{\rm min}$ and
$\theta_{\rm max}$, supplemented by the effects of $\lambda$.

\section{Finding Fiducial Models: Basic Approach}
\label{sec:approach}

Our primary objective is to constrain the physical properties of the
wind. The empirical wind models described in the previous section
involve a rather large number of parameters, too many for it to be
computationally practical to compute large grids of model spectra that
explore the complete parameter space. Therefore, we have chosen to
accept previous determinations of the system parameters, where
possible, and focus our study on the major parameters of the wind
itself. 

The distance to RW Tri has been determined to be $D = 341^{-31}_{+38}$
pc from its parallax \citep{McArthur1999}.  This is almost identical
to the distance obtained by \cite{Rutten1992} from the analysis of the
optical eclipse light curve.  According to \cite{Rutten1992} the WD mass for RW Tri is about
0.7\MSOL, which is typical of WDs in other systems and the mass ratio
is fairly close to 1.  In their study they also estimate that the inclination
for RW Tri is about 75$\degr$, although this is somewhat uncertain and
values as low as 67$\degr$ \citep{Kaitchuck1983} and as high as
80$\degr$ \citep{Mason1997} have been
suggested. From spectroscopic studies, \cite{Kaitchuck1983} also determined the $K$ velocity of the WD. The amplitude of this velocity is roughly 200~$\VEL$. Table \ref{tab:system_pars} lists the system parameters we
adopted for RW Tri. 
\begin{table}[t]
  \centering
  \caption{System Parameters for RW~Tri and UX~UMa.}
  \begin{tabular}{@{}l l@{} @{}l l@{}}
    \hline\hline
    \multicolumn{1}{c}{System} & \multicolumn{1}{c}{Parameter} & \multicolumn{1}{c}{Value} & \multicolumn{1}{c}{Reference} \\ 
     \hline
    RW~Tri  & $M_{\rm WD}$ & $0.7 $~$M_{\odot}$ & \cite{Rutten1992}\\
     & $M_{2}$ & $0.6 $~$M_{\odot}$ & \cite{Rutten1992}\\
     & $i$ & $75\degr$ & \cite{Rutten1992}\\
     & $D$ & $330$~pc & \cite{Rutten1992}\\
     & $\dot M_{\rm acc}$ & $\sim 10^{-8.0} $~$M_{\odot} $~yr$^{-1}$ & \cite{Rutten1992}\\
     & $K_1$ & $\sim200~\VEL$ & \cite{Kaitchuck1983}\\
    \hline
    UX~UMa & $M_{\rm WD}$ & $0.47 \pm 0.07 $~$M_{\odot}$ & \cite{Baptista1995}\\
           &          & $0.78 \pm 0.13 $~$M_{\odot}$ & \cite{VandePutte2003}\\
     & $M_{2}$ & $0.47 \pm 0.10 $~$M_{\odot}$& \cite{Baptista1995}\\
     &           &  $0.47 \pm 0.07$~$M_{\odot}$& \cite{VandePutte2003}\\
     & $i$ & $71.0 \pm 0\degr$\hspace {-1.25mm}$.6$& \cite{Baptista1995}\\
     & $D$ & $345 \pm 34$ pc& \cite{Baptista1995}\\
     & $\dot M_{\rm acc}$ & $10^{-8.0 \pm 0.2} $~$M_{\odot} $~yr$^{-1}$& \cite{Baptista1995}\\
     & $K_1$ & $\sim 200~\VEL$ & \cite{Schlegel1983}\\
    \hline

  \end{tabular}
\label{tab:system_pars}
\end{table}

The main source for the continuum flux in nova-like variables is
normally the accretion disk. For a given wind model, the ionization
structure is primarily determined by the radiation temperature $T_{\rm R}$,
which, for a steady state, is controlled by the effective
temperature ($T_{\rm eff}$) of the disk.
Assuming that the disk radiates as an
ensemble of stellar atmospheres, as described by
Equation (\ref{eq:effective_temp}), then the mass accretion rate is a
function of the observed flux, the distance, inclination, and the mass
and radius of the WD. Thus, we can use the observed continuum flux and system
parameters to constrain the mass accretion rate and avoid treating it
as a completely free parameter in our studies. For this exercise, we
restricted the wavelength range for the continuum fits to the regions
between the \ion{Si}{4}, \ion{C}{4}, and \ion{He}{2} lines which show a
clean continuum in the observations (in contrast to the complex FUV
region). For RW Tri, the continuum fits resulted in an accretion rate
of \EXPU{8.3}{-9}{\MSOL~yr^{-1}}  (see
Figure \ref{fig:rwtri_obs_spec}). Accounting for the uncertainties in
the inclination ($75^{+5}_{-8} ~\degr$), the distance ($341^{+38}_{-31}$ pc)
and the WD mass (here, an uncertainty of one-tenth of the solar mass
was assumed), the valid range for the accretion rate would
be \EXPU{2.7}{-9}{\MSOL~yr^{-1}} to  \EXPU{2.6}{-8}{\MSOL~yr^{-1}},
corresponding to a maximum effective temperature of the disk between
$T_{\rm eff} = 4.9 \times 10^4$~K
and $6.9 \times 10^4$~K.

The most determined effort to establish the system parameters for UX
UMa was made by \cite{Baptista1995}. They used a (different) set of
high time resolution, low spectral resolution \HST\ spectra, and modeled
the time-dependent  contribution of the disk and WD to the continuum
to identify the phases associated with the eclipse of the WD.   From
this, they inferred a WD mass of $0.47\pm0.07 \MSOL$, and a mass
ratio ($M_2/M_{\rm WD}$) of 1$\pm$0.1 and an inclination of $71\pm1\degr$.
They then derived a distance of 345$\pm$34 pc from eclipse maps of the
system obtained from the combination of UV and $R$-band light
curves. Since then, most workers, including most
recently \cite{Linnell2008}, have adopted these parameters for UX
UMa. There are various reasons to question the accuracy of the
measurements. First, if correct, the WD mass in UX UMa is much lower
than that of most WDs (including most WDs in CV systems), and the mass ratio is
at the limit allowed by conservative mass transfer. Second,  at the
temperature determined by \cite{Baptista1995} for the WD, the WD
should have been evident in the \FUSE\ spectra of UX UMa,
but \cite{Froning2003}, analyzing time-resolved \FUSE\ observations of
UX UMa, were unable to identify the WD ingress and egress in the light
curve. Despite these concerns, we too adopt the parameters
of \cite{Baptista1995}, since no alternative values have been proposed
with greater credibility.  They are also listed in
Table \ref{tab:system_pars}. The value for the $K$ velocity of UX UMa, again $\sim 200$~$\VEL$ was adopted from \cite{Schlegel1983}.

According to the continuum fit (see Figure \ref{fig:uxuma_obs_spec}),
the accretion rate for UX UMa is a little higher than RW Tri. It was
determined as  \EXPU{1.2}{-8}{\MSOL~yr^{-1}}  but values within the
range of \EXPU{6.9}{-9}{\MSOL~yr^{-1}}
to \EXPU{1.9}{-8}{\MSOL~yr^{-1}} would still be consistent with the
system parameters considering the uncertainties in the WD mass, the
inclination, and distance. This corresponds to a maximum disk effective temperature
range of $T_{\rm eff} = 4.6 \times 10^4$~K to $4.9 \times 10^4$~K.

In order to study the wind geometry and the ionization structure of
the mass outflow in UX UMa and RW Tri, we used {\sc python} to
develop ``point'' models for both systems. To avoid 
the computationally demanding brute-force method of computing grids of
models and searching for the best-match spectrum, we adopted a
hierarchical approach to searching the wind parameters. 
Our basic methodology was to come as close as possible to the
observations by adjusting the overall ionization structure in the wind
first. For that we started with the KWD95 descriptions and the wind
parameters derived by \cite{Knigge1997},
applied to both our subjects of study, RW Tri and UX~UMa,
and explored different values of the mass
accretion and the mass-loss rate --- these are the primary parameters
that determine the ionization conditions. However, we restricted the
mass accretion rate to the values compatible with the results of disk
continuum fits (see above). After exploring the $\dot 
M_{\rm wind} \times \dot M_{\rm acc}$ parameter space, the influence of each
of the remaining free wind parameters on the spectrum was tested,
allowing us to successively obtain closer agreement with the
observations. When using SV93 point models, the same approach was
used except that the $\dot M_{\rm wind} \times \dot M_{\rm acc}$ space was not
re-explored and wind parameters which roughly correspond to the best
KWD95 model were used as a starting point. With this approach we were able to
explore the sensitivity of the spectrum to the main parameters although it
did not lead to an exhaustive investigation of the complete parameter space.
In general, determining our ``best" models involved subjective criteria since 
perfect matches were never obtained. In these choices, we prioritized the overall
strength and shape of the major emission lines.

Applying these strategies, we were able to find good point models for
both systems using both wind treatments, SV93 and KWD95. The following
sections are dedicated to presenting these models (both their
successes and failures) and the conclusions drawn from them.
Throughout our discussion, we focus on the three prominent emission
lines of \ion{N}{5}, \ion{Si}{4}, and \ion{C}{4}. As we shall show,
these lines form across a wide range of conditions in the wind and,
together, probe a significant fraction of the outflowing material. In
addition, as resonance lines in Li- and Na-like ions, they involve
relatively simple atomic physics which should be well described by our
radiative transfer simulations. We will, however, also comment on
other spectroscopic features formed by lower ionization state metal
ions (which exist predominantly in the coolest/densest parts of the
models) and recombination line features of both H and He. 

\section{Results}
\label{sec:results}

We now present our point models for both RW~Tri and UX~UMa (see below)
and then discuss the similarities and differences we infer for the wind
structures in these systems.

Since we allowed for some flexibility in the mass accretion rate for a
given set of system parameters, we have also allowed some freedom in
the distance we adopt for both objects. Specifically, we always
renormalized the model spectra to the observations in the discussion
below.  For this renormalization the continua of model and data were
compared between \ion{Si}{4} and \ion{C}{4}. However, 
the implied deviations from the
distances given in Table~\ref{tab:system_pars} are well within the uncertainties
since the point models for both systems have accretion rates very
close to the values determined by our disk continuum fits. We also corrected all pre-eclipse model spectra for the $K$ velocity (redshift) of the WD. This shift however was rather small --- from $0.3$~\AA\ at the blue end to $0.6$~\AA\ at the red end of the spectrum. 

\subsection{RW Tri}

Table~\ref{tab:kwd_pars} gives the parameters of our best model for RW
Tri obtained with the KWD95 wind description via our strategy
described in Section~\ref{sec:approach}. 
Exploration of the $\dot M_{\rm wind} \times \dot M_{\rm acc}$ space for this
object led us to prefer $\dot M_{\rm acc} \simeq 8 \times
10^{-9}$~$M_{\odot}$~yr$^{-1}$ combined with $\dot M_{\rm wind} \simeq
4.8 \times 10^{-10}$~$M_{\odot}$~yr$^{-1}$ ($\sim 0.06 \dot{M}_{\rm acc}$). This best value of $\dot M_{\rm acc}$ is comfortably within the
allowed range discussed in Section~\ref{sec:approach} and implies a
distance to RW~Tri of $D = 350$~pc which is
reasonable.
We found that a low value of the mass-loss exponent $\alpha$ was
required to achieve an acceptable match to the relative line strengths
in this system. A low value of $\alpha$ means that the density near the
rotational axis of the system is small, allowing for a mean higher
ionization state in this region. This was required in order to produce
a sufficiently strong \ion{N}{5} emission line, relative
to \ion{C}{4}. Our exploration of parameters indicated that the
collimation of the RW Tri wind is not very strongly constrained by the
spectra. Only very low collimations (low values of $d \lesssim 10$) can clearly be
ruled out since they lead to clear P-Cygni profiles in \ion{C}{4}
which are not observed. In our best model, a slightly higher
collimation was adopted than in \cite{Knigge1997}, since we found that
it led to a slightly better match with the data. 
\begin{table}[t]
  \tiny
  \centering
  \caption{Parameters for Our Best KWD95 Model for RW~Tri and UX~UMa.}
  \begin{tabular}{@{}l @{}l @{}l @{}l@{}}
    \hline\hline
    \multicolumn{1}{c}{Parameter} & \multicolumn{1}{c}{RW Tri} & \multicolumn{2}{c}{UX UMa} \\
    \hline
    & \multicolumn{1}{c}{Point Model} & \multicolumn{1}{c}{Point Model} & \multicolumn{1}{c}{KD 97~$^{\rm a}$} \\
    \hline
    $M_{\rm WD}$ &  $0.7 $~$M_{\odot}$& $0.47 $~$M_{\odot}$ &  \\ 
    $M_{2}$ &  $0.6 $~$M_{\odot}$& $0.47 $~$M_{\odot}$ & \\
    $R_{\rm WD}$ & \EXPU{8.0}{8}{cm}& \EXPU{9.7}{8}{cm}& \\
    $R_{\rm disk}$ & \EXPU{2.4}{10}{cm}& \EXPU{2.9}{10}{cm}& \\
    $i$ & $75\degr$\hspace{-1.2mm}$.0$ & $71\degr$\hspace{-1.2mm}$.0$ & \\
    $D$~$^{\rm b}$ & 349~pc & 374~pc & \\
    $\dot M_{\rm disk}$ & \EXPU{8.0}{-9}{\MSOL ~yr^{-1} } & \EXPU{1.3}{-8}{\MSOL ~yr^{-1}} & \EXPU{2.7}{-8}{\MSOL ~yr^{-1}}\\
    $T_{\rm eff}$~$^{\rm c}$ & \EXPU{5.7}{4}{K}  &\EXPU{5.1}{4}{K} & \\ 
    $\dot M_{\rm wind}$ &  \EXPU{4.8}{-10}{\MSOL ~yr^{-1} } &\EXPU{1.0}{-9}{\MSOL ~yr^{-1}}   & \EXPU{1.0}{-9}{\MSOL ~yr^{-1}}  \\
    $d$ & $20$& $35$ & $15$\\
    $\alpha$ & $0.1$& $0.5$ & $0.5$\\
    $f$ & $3.0$& $3.0$ & $3.0$\\
    $R_{\rm s}$ &\EXPU{4.0}{10}{cm} & \EXPU{4.9}{10}{cm}  & \EXPU{4.9}{10}{cm}\\
    $\beta$ & $4.5$ & $1.0$ & $4.5$\\
    \hline
  \end{tabular}\\
{
  \footnotesize
  \begin{flushleft}
    {\bf Notes.}\\
    $^{\rm a}$ {\footnotesize System and wind parameters from \cite{Knigge1997}. These parameters were used as a starting point when deriving our point models.}\\
    $^{\rm b}$ {\footnotesize Distance implied by the renormalization of the model continuum.}\\
    $^{\rm c}$ {\footnotesize Maximum effective temperature in the disk determined via
      Equation (\ref{eq:effective_temp}) from $M_{\rm WD}$, $R_{\rm WD}$, $\dot M_{\rm disk}$.}\\
  \end{flushleft}
}
\label{tab:kwd_pars}
\end{table}

\begin{figure}[t]
  \centering
  \ifthenelse{\boolean{OnlineEdition}}{
    \includegraphics[width=0.48\textwidth]{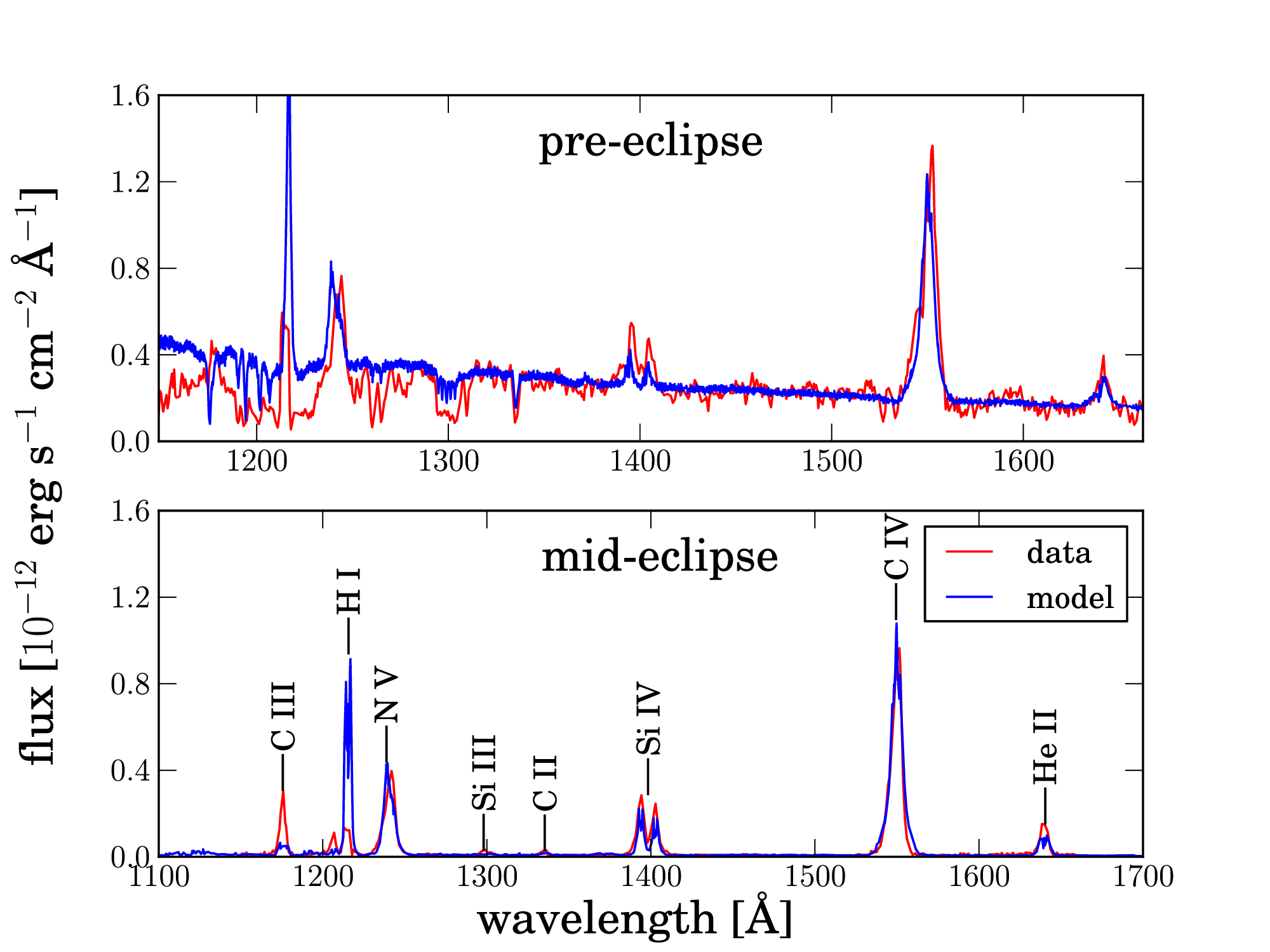}
  }
  {
    \includegraphics[width=0.48\textwidth]{rwtri_kwd95_spec_1_bw.eps}
  }
  \caption{Comparison between RW Tri \HST\ observations and our best KWD95 model.}
  \label{fig:rwtri_kwd_spec}
\end{figure}
\begin{figure}[t]
  \centering
  \ifthenelse{\boolean{OnlineEdition}}{
    \includegraphics[width=0.48\textwidth]{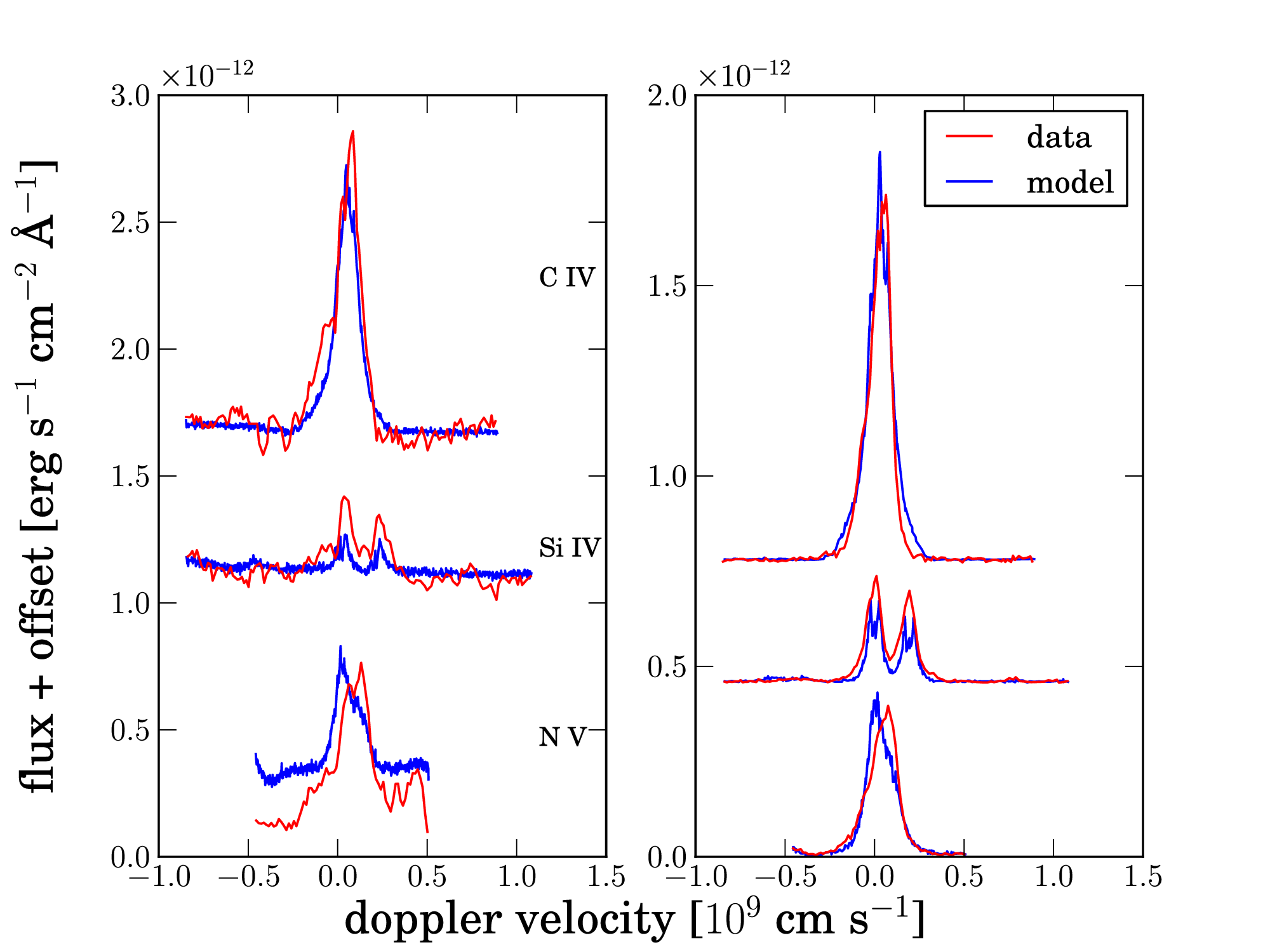}
  }
  {
    \includegraphics[width=0.48\textwidth]{rwtri_kwd95_lines_1_bw.eps}
  }
  \caption{Detailed comparison of line profiles for RW Tri and our KWD95 model during pre-eclipse (left) and mid-eclipse (right).}
  \label{fig:rwtri_kwd_profiles}
\end{figure}
\begin{figure}[t]
  \centering
  \includegraphics[width=0.48\textwidth]{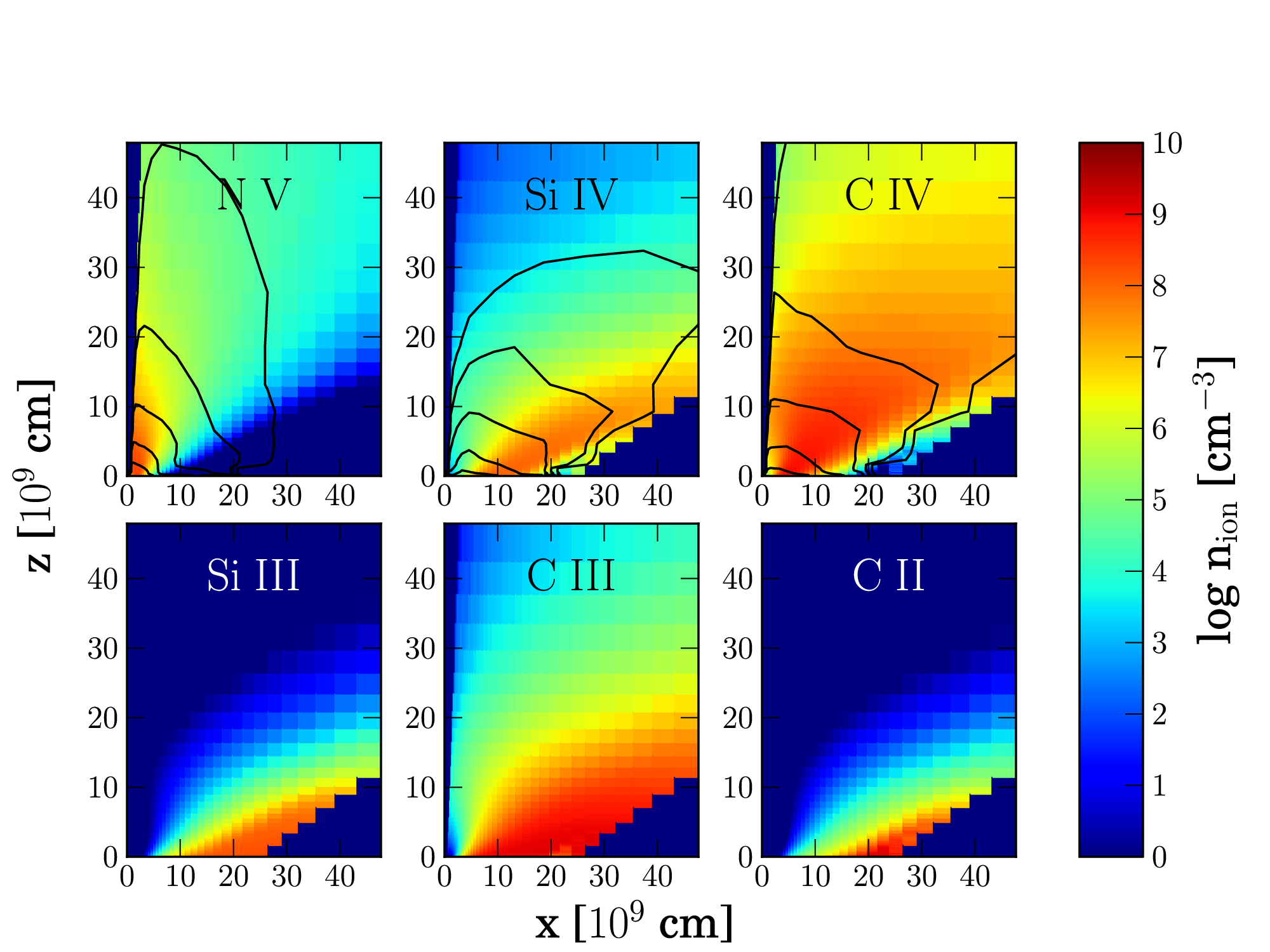}
  \caption{Number densities for various ions as a function of position in
  our wind model for RW~Tri (adopting the KWD95 prescription). 
  Only the inner portion
  of the wind that produces substantial scattering is shown (our
  computational grid extended to \POW{13}{cm}). The black contour
  lines indicate the line-forming regions of the main three resonance
  lines (upper panels). Specifically, they are contours of 
  the number of relevant last scattering events in a
  computational grid cell normalized
  by the cell volume. The four contours are drawn for one decade intervals in
  this quantity which, as expected, decreases outward.
}
  \label{fig:rwtri_kwd_ions}
\end{figure}
\begin{figure}[t]
  \centering
  \includegraphics[width=0.48\textwidth]{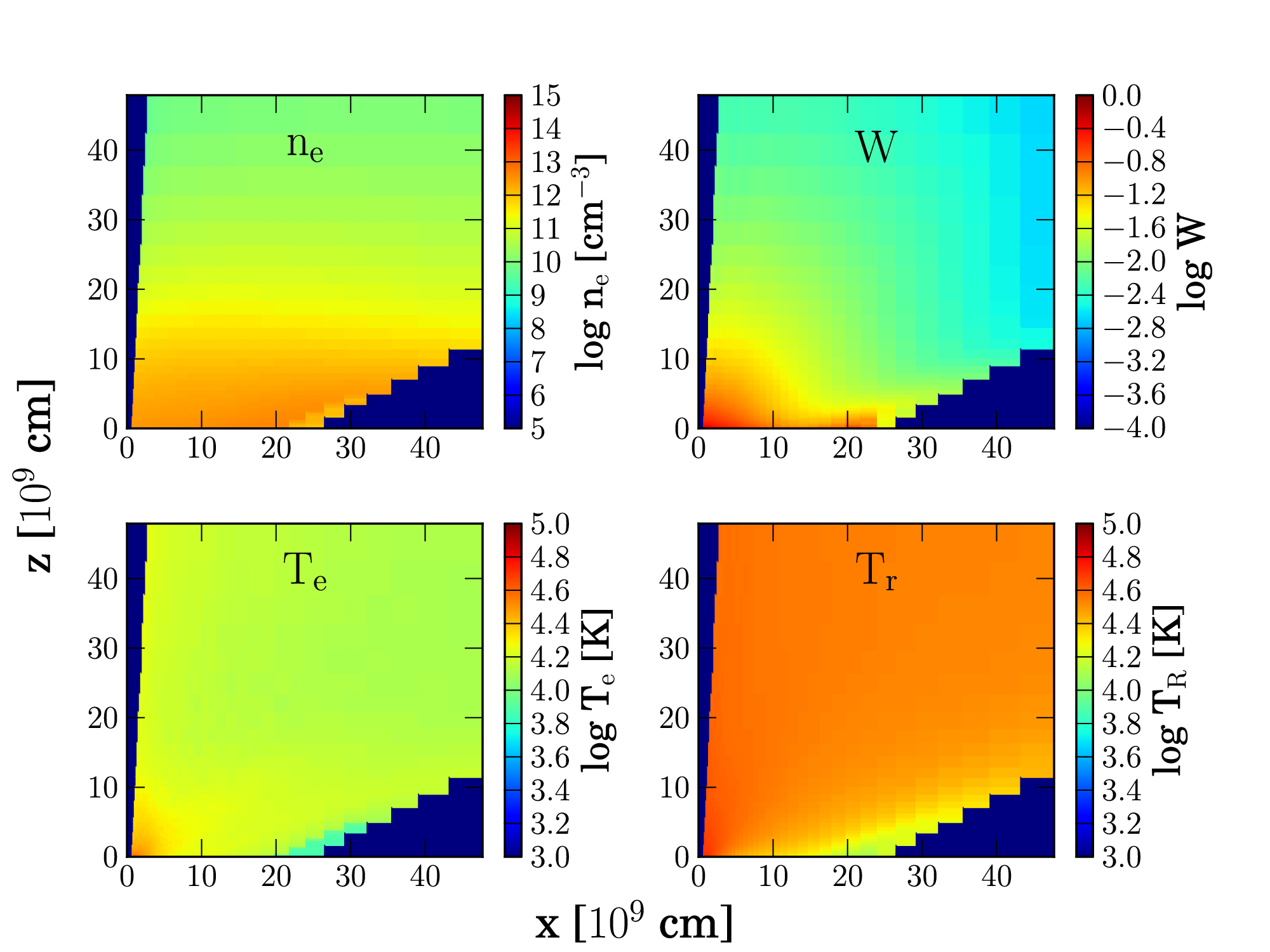}
  \caption{Temperature, dilution factor, and density structure of the RW Tri KWD95 model.}
  \label{fig:rwtri_kwd_temp}
\end{figure}

The spectrum obtained for this model at an inclination angle of $75\degr$ is compared to the observations in Figure~\ref{fig:rwtri_kwd_spec}. Details of the line profiles for the major resonance transitions of \ion{N}{5}, \ion{Si}{4}, and \ion{C}{4} are shown in Figure~\ref{fig:rwtri_kwd_profiles}.
These are in good agreement with the observations with
regard to both line width and strength. Since they form under
different ionization conditions, they are present in quite different
regions of the wind; see Figure~\ref{fig:rwtri_kwd_ions}. In
particular, \ion{N}{5} probes the highest ionization state material
which exists in the innermost streamlines where the radiation
temperature is high owing to the abundance of ultraviolet photons
radiated by the inner regions of the disk (see
Figure~\ref{fig:rwtri_kwd_temp}). In contrast, \ion{Si}{4} is confined
to a relatively narrow equatorial band lying across the base of flow
streamlines that rise further out in the disk. This region has lower $T_{\rm R}$
since it is exposed to the radiation field from the cooler outer disk regions. 
This, combined with the high electron densities ($n_{\rm e}$) at the base of the 
flow, allows relatively low-ionization state material to dominate this region. 
\ion{C}{4} is present across a much wider region of the wind than either 
\ion{N}{5} or \ion{Si}{4}, occupying most of the space between the polar 
regions (rich in \ion{N}{5}) and the equatorial band of \ion{Si}{4}. \ion{C}{4}
also exists throughout much of the outer wind allowing for a significantly 
extended region of line formation. This is a consequence of the relatively 
uniform distribution of $T_{\rm R}$ and $n_{\rm e}$ in the outer parts of the flow
(see Figure \ref{fig:rwtri_kwd_temp}).

In Figure \ref{fig:rwtri_kwd_ions}, contours are drawn to indicate the
line-forming regions. Specifically, these contour lines indicate where
the last scattering events for Monte Carlo quanta which
contribute to specific resonance lines took place. As one would expect
from the ion distribution, \ion{C}{4} has a somewhat extended region
of line formation, while \ion{N}{5} and \ion{Si}{4} are more confined
to the inner parts of the flow, being concentrated in the polar or more 
equatorial regions, respectively. 
Given these significant differences in region of line formation, it is
a powerful confirmation of the wind model that the relative strengths
and shapes of these main line features can be reproduced, both in and out
of eclipse, by a single model in which the ionization
conditions are controlled by the radiation produced by the accretion disk.

In addition to the three major lines, several metal lines associated
with lower ionization state material appear in the spectra. In the
pre-eclipse spectrum, absorption by \CiiLxiii\ and \SIiiiL\ are present in both
model and observed spectra although \ion{Si}{3} is rather too weak in the model. \CiiiLxi\ also 
appears in both model and observation but its character is less well represented
--- in particular it appears as absorption in the model pre-eclipse spectrum but
emission in the data. During eclipse this feature is in emission in the model
but it is too weak. Taken together, this suggests that the wind model may place
\ion{C}{3} too close to the disk plane --- a better match might be obtained if 
more of this ion existed in the extended (uneclipsed) wind region with less 
lying along our relatively high-inclination line of sight. Recombination line 
emission in both \ion{H}{1} \LA\ $\lambda1215$ and \HEiiL\ is also predicted by
the model. For \ion{He}{2}, the recombination line strength is too small in the
model but for \ion{H}{1} Ly$\alpha$ it is significantly too strong in the
pre-eclipse spectrum. This line forms rather deep in the wind models, meaning 
that its emission line flux is significantly affected during eclipse (compare 
model line strengths in Figure~\ref{fig:rwtri_kwd_spec}). It is therefore much 
more sensitive to the structure at the base of the wind than the other strong 
emission lines and may indicate a shortcoming of our wind models in describing 
those regions. 

\begin{table}[t]
  \centering
  \caption{Parameters for Our Best SV93 Models for RW~Tri and UX~UMa.}
  \begin{tabular}{ @{}l l l@{} }
    \hline\hline
    \multicolumn{1}{c}{Parameter} & \multicolumn{1}{c}{RW Tri} & \multicolumn{1}{c}{UX UMa} \\
    \hline
    $M_{\rm WD}$ &  0.7~\MSOL& 0.47~\MSOL\\
    $M_{2}$ &  0.6~\MSOL& 0.47~\MSOL\\
    $R_{\rm WD}$ & \EXPU{8.0}{8}{cm} &\EXPU{9.7}{8}{cm}\\
    $R_{\rm disk}$ & \EXPU{2.4}{10}{cm} & \EXPU{2.9}{10}{cm}\\
    $i$ & $75\degr$\hspace{-1.2mm}$.0$ & $71\degr$\hspace{-1.2mm}$.0$\\
    $D$~$^{\rm a}$ & 345~pc & 374~pc \\
    $\dot M_{\rm disk}$ & \EXPU{8.0}{-9}{\MSOL~yr^{-1}}& \EXPU{1.3}{-8}{\MSOL~yr^{-1}} \\
    $T_{\rm eff}$~$^{\rm b}$ & \EXPU{5.7}{4}{K} & \EXPU{5.1}{4}{K}\\ 
    $\dot M_{\rm wind}$ & \EXPU{4.8}{-10}{\MSOL~yr^{-1}} &  \EXPU{1.0}{-9}{\MSOL~yr^{-1}}\\
    $R_{\rm min}$ & $R_{\rm WD}$& $R_{\rm WD}$\\
    $R_{\rm max}$ & $R_{\rm disk}$& $R_{\rm disk}$\\
    $\theta_{\rm min}$ & $2\degr$\hspace{-1.2mm}$.9$ & $1\degr$\hspace{-1.2mm}$.4$\\
    $\theta_{\rm max}$ & $56\degr$\hspace{-1.2mm}$.3$ & $36\degr$\hspace{-1.2mm}$.9$\\
    $\gamma$ & $1$ & $1$ \\
    $\lambda$ & $0.6$ & $-0.5$\\
    $f$ & $3$& $4$\\
    $R_{\rm s}$ &\EXPU{2.0}{11}{cm} & \EXPU{4.9}{10}{cm} \\
    $\alpha$ & $2.2$ & $1.0$\\
    \hline
  \end{tabular}\\
  {
    \begin{flushleft}
      {\bf Notes.}\\
      $^{\rm a}$ {\footnotesize Distance implied by the renormalization of the model continuum.}\\
      $^{\rm b}$ {\footnotesize Maximum effective temperature in the disk determined via Equation (\ref{eq:effective_temp}) from $M_{\rm WD}$, $R_{\rm WD}$, $\dot M_{\rm disk}$.}\\
    \end{flushleft}
  }
  \label{tab:sv_pars}
\end{table}

\begin{figure}[t]
  \centering
  \ifthenelse{\boolean{OnlineEdition}}{
    \includegraphics[width=0.48\textwidth]{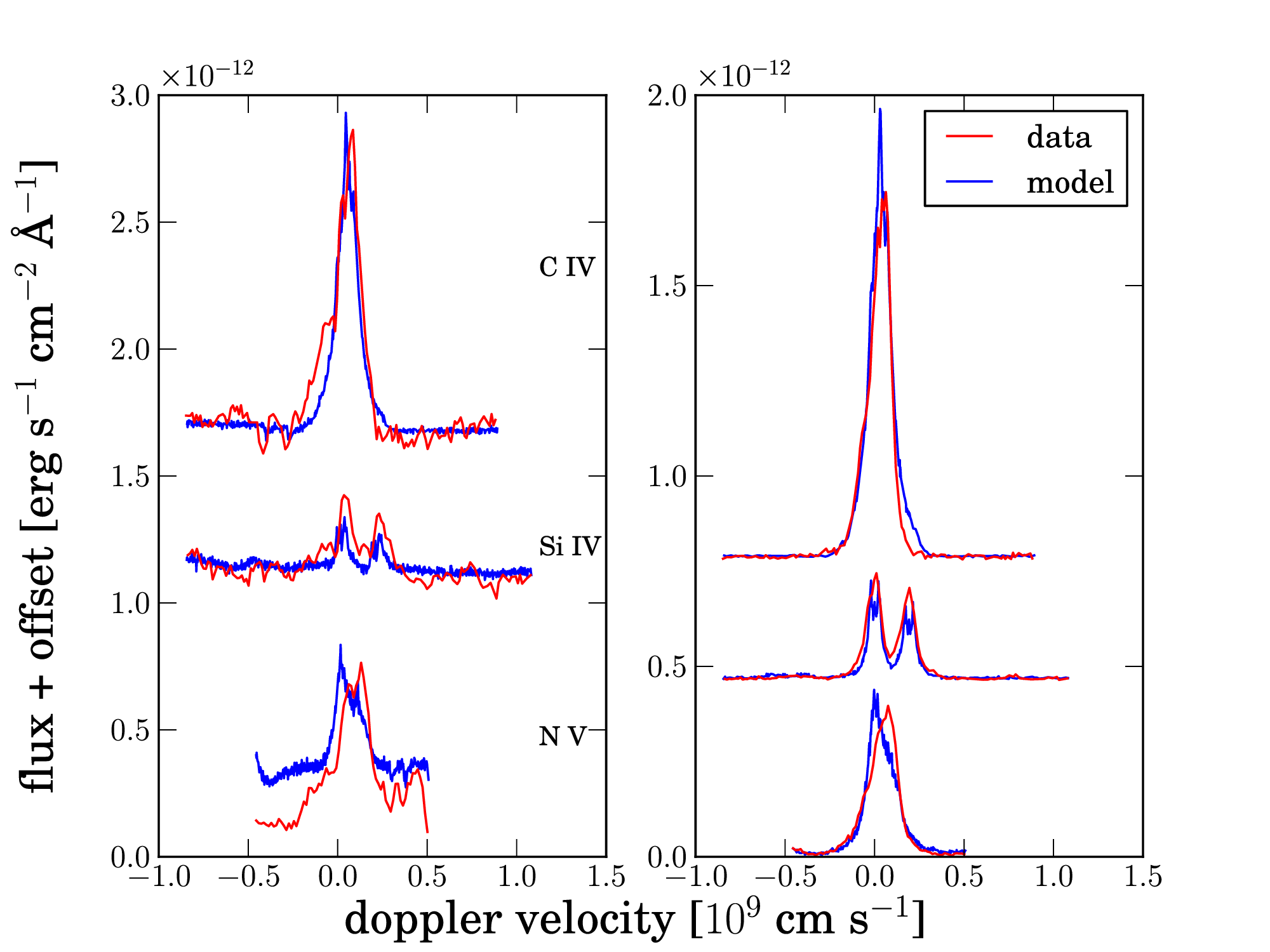}
  }
  {
    \includegraphics[width=0.48\textwidth]{rwtri_sv93_lines_1_bw.eps}
  }
  \caption{Details of line profiles for the RW Tri SV93 model during pre-eclipse (left) and mid-eclipse (right).}
  \label{fig:rwtri_sv_profiles}
\end{figure}

To establish the sensitivity of our conclusions about the line formation 
regions to the choice of how the wind is parameterized, we have also developed a
wind model for RW~Tri using the SV93 wind prescription. The parameters for this
model are given in Table~\ref{tab:sv_pars} and its spectrum is shown in Figure 
\ref{fig:rwtri_sv_profiles}. These show comparably good agreement in the 
spectral features as found with the KWD95 model. Moreover, despite the difference
in parameterization, they indicate characteristically similar geometries for the 
distribution of ionization states  as found with the KWD95 wind prescription. 

\subsection{UX UMa}

We employed a similar methodology to obtain point models for UX UMa with both 
the KWD95 and SV93 wind prescriptions. The parameters of these models are given
in Tables~\ref{tab:kwd_pars} and \ref{tab:sv_pars}. 

\begin{figure}[t]
  \centering
  \ifthenelse{\boolean{OnlineEdition}}{
    \includegraphics[width=0.48\textwidth]{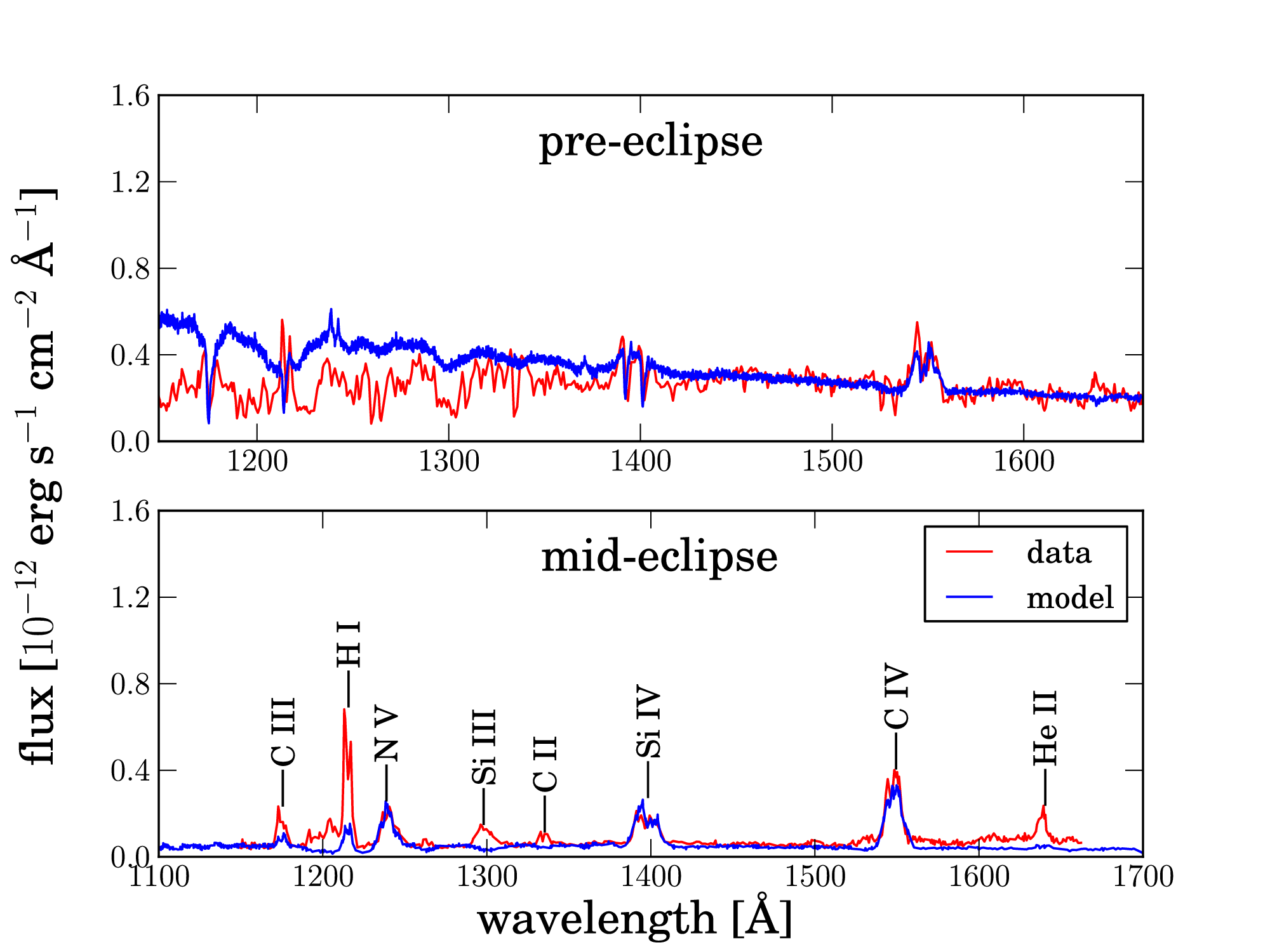}
  }
  {
    \includegraphics[width=0.48\textwidth]{uxuma_kwd95_spec_1_bw.eps}
  }
  \caption{Comparison between UX UMa \HST\ observations and our best KWD95 model.}
  \label{fig:uxuma_kwd_spec}
\end{figure}

\begin{figure}[t]
  \centering
  \ifthenelse{\boolean{OnlineEdition}}{
    \includegraphics[width=0.48\textwidth]{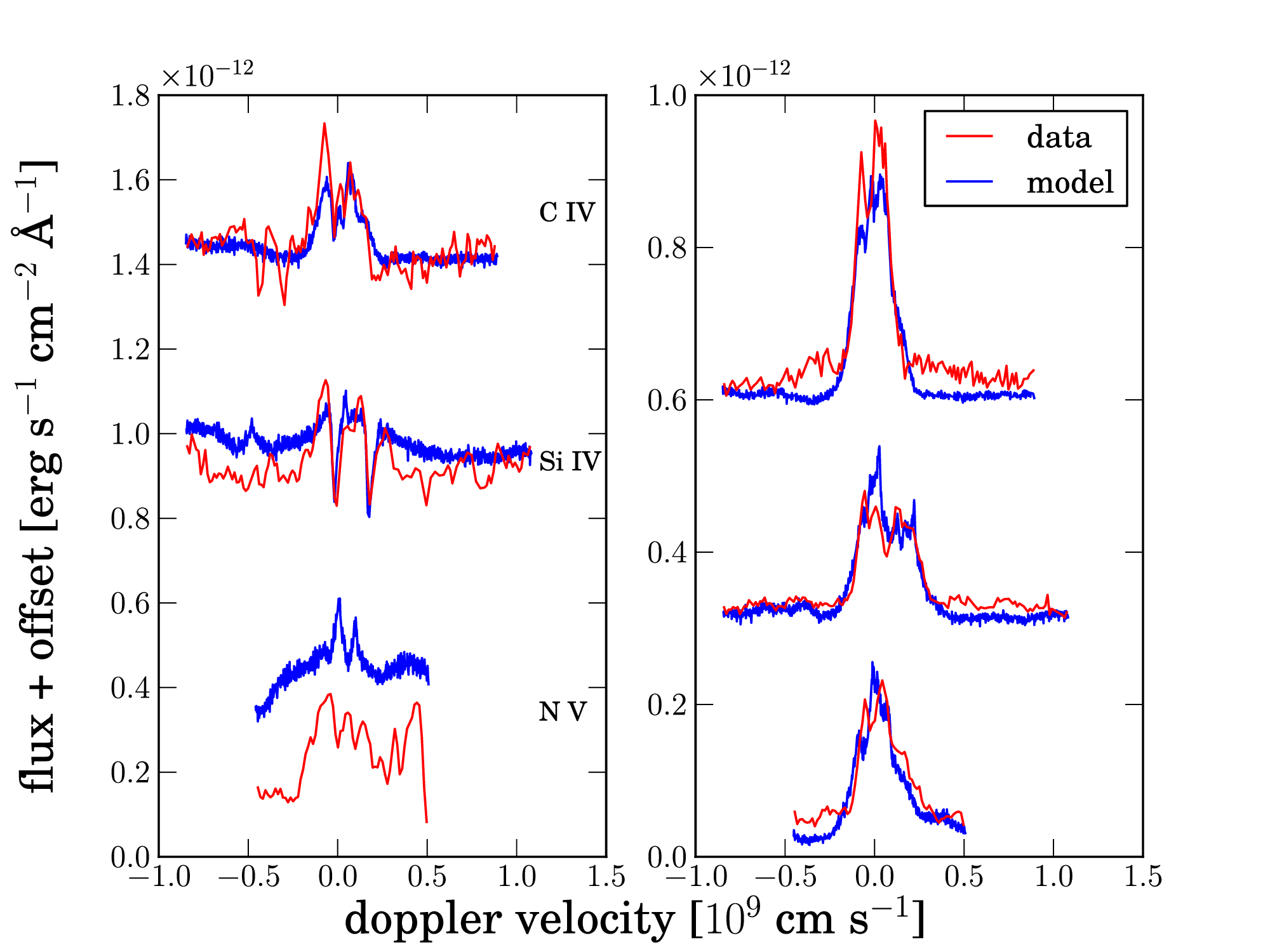}
  }
  {
    \includegraphics[width=0.48\textwidth]{uxuma_kwd95_lines_1_bw.eps}
  }
  \caption{Details of line profiles for the UX UMa KWD95 model during pre-eclipse (left) and mid-eclipse (right).}
  \label{fig:uxuma_kwd_profiles}
\end{figure}

Based on the fits to the continuum, we require a lower effective
temperature for UX~UMa meaning that the  inner disk radiation is
redder than in RW Tri. Thanks to this, a good match to the relative
strengths of  emission in \ion{N}{5}, \ion{C}{4} and \ion{Si}{4} was
obtained without having to adopt a low value for $\alpha$
(cf.\ discussion of \ion{N}{5} in RW~Tri above). An important
difference in the observed spectra of UX~UMa compared to RW~Tri,
however, is the appearance of narrow absorption dips in
the \ion{Si}{4} profile out of eclipse. We could not reproduce these dips via 
simple changes in either $\dot{M}_{\rm wind}$ or $\alpha$. To explain these 
features, it is necessary to introduce \ion{Si}{4} ions along the observer's 
line of sight which requires a more equatorially concentrated \ion{Si}{4} 
population. By exploring the velocity-law parameters, we found that this could 
readily be explained by a more-rapidly accelerating flow (specifically, we 
reduce the acceleration exponent $\beta$; qualitatively similar results can be 
obtained with a smaller acceleration length $R_{\rm s}$). A more rapid 
acceleration reduces the density allowing \ion{Si}{4} to replace \ion{Si}{3} 
around the base of the wind. Together with a rather high collimation of the 
wind, which was necessary to ensure the right width and blueshift of the 
\ion{Si}{4} absorption dips, this approach led to a reasonable match of the 
three main line profiles (see Figures~\ref{fig:uxuma_kwd_spec} and 
\ref{fig:uxuma_kwd_profiles}).
Our low value of $\beta = 1.0$ is the most substantial difference in 
the wind parameters between this study and that of \cite{Knigge1997}: 
they favored a significantly more gradual acceleration ($\beta =
4.5$) but modeled only the C~{\sc iv} profile and did not
compute the wind ionization state in detail. Admittedly, they obtained
a somewhat better fit to the C~{\sc iv} profile than we have (see
their Figure~4); however, we find that the additional constraints on
the wind properties by the complex Si~{\sc iv} line profile require a
modification to the density structure around the base of the wind
which are consistent with much more rapid acceleration.
A negative consequence of the change in the density structure around the base 
of the wind is that this model significantly underpredicts the strengths of 
features associated with the lowest ionization material (e.g., \ion{C}{2} and 
\ion{Si}{3} wind features are entirely absent from the model spectra, while the
recombination \ion{H}{1} \LA\ and \HEiiL\ lines are too weak). 

While the major emission lines in the model match the data well, there
is a discrepancy between the slope of the model and the observed disk continuum
in the pre-eclipse state. The model continuum, which reflects that of a 
steady-state disk, is steeper than the observed spectrum, a fact that has been 
known about UX UMa for some time \cite[see, e.g.][]{Knigge1998}. One 
possibility is that the run of temperatures in the disk simply departs from 
that of a steady-state disk, an idea that was explored in detail by 
\cite{Linnell2008} without, in our opinion, finding an obviously better model 
for the disk emission. Our models would suggest that, despite the apparent 
mismatch in the continuum slope, the ionization conditions in the wind required
to produce the correct relative strengths of the emission lines can be obtained
from a standard, steady-state disk configuration: if we were to significantly
alter the distribution of $T_{\rm eff}$ for the disk in our fiducial model, the ionization conditions (and therefore line emission) in the wind would change. 
Although this might be accounted for by modifications to other wind parameters,
on balance we favor an alternative possibility, namely, that the mismatch of the 
continuum slope is the result of occultation of the inner portion of the disk
along our line of sight.  This is certainly not an established fact,
but UX UMa is known to exhibit pre-eclipse dips in the UV that are
attributed to this effect; it is possible that portions of the disk are
occulted all of the time, as is known to be the case in some SW Sex
stars, including DW UMa \citep{Knigge2004}.  It seems less likely to
us that the disk is truncated because this would necessarily reduce
the peak value of $T_{\rm eff}$ in the disk and therefore predict a less
ionized wind.  

\begin{figure}[t]
  \centering
  \includegraphics[width=0.48\textwidth]{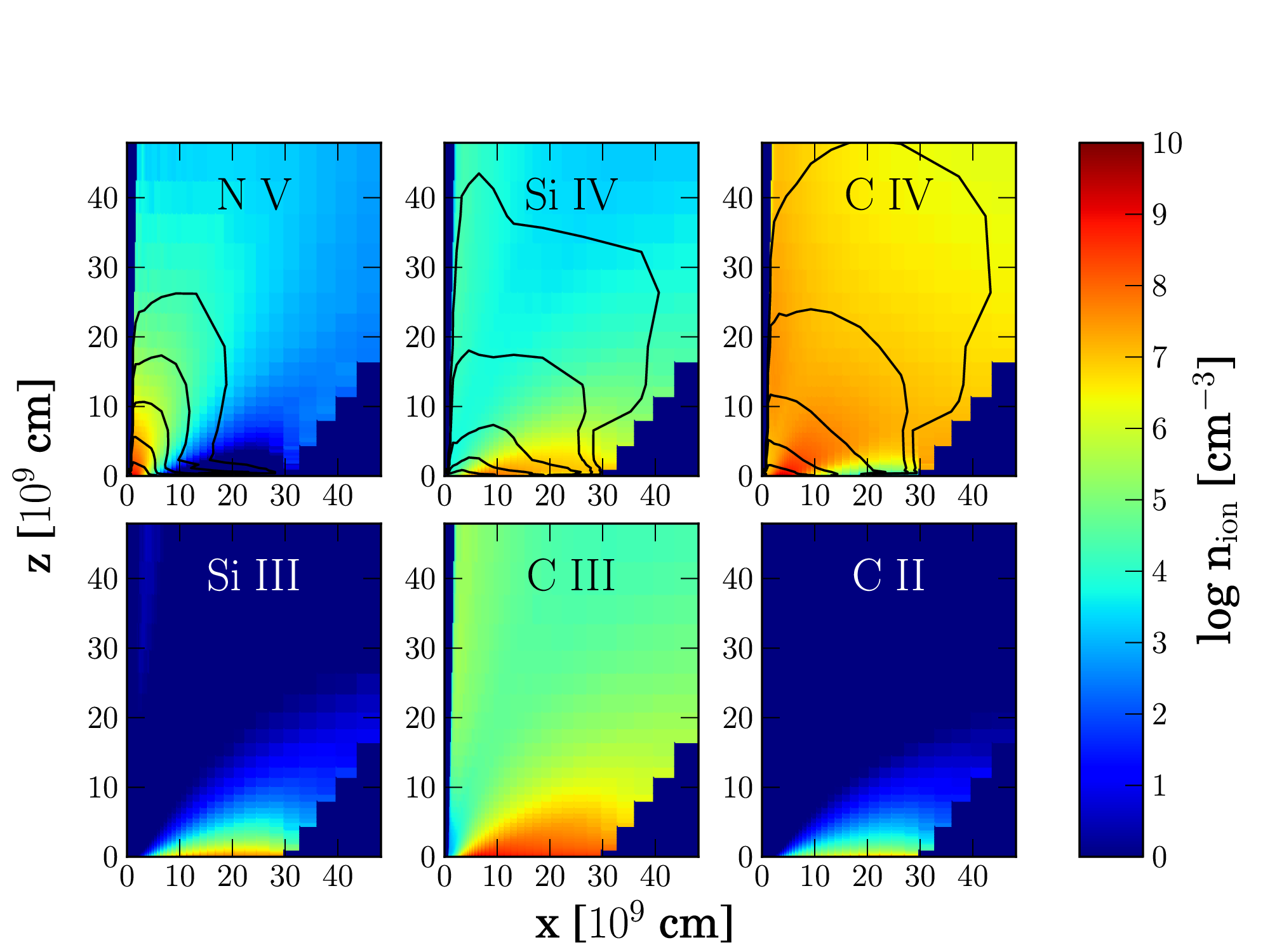}
  \caption{Ionization structure of the UX UMa KWD95 model. Again the line-forming regions are shown by contour lines (cf.\ Figure \ref{fig:rwtri_kwd_ions}).}
  \label{fig:uxuma_kwd_ions}
\end{figure}

The ionization structure of the UX UMa wind model (shown in Figure~
\ref{fig:uxuma_kwd_ions}) is different in detail but similar in character to 
that in the models fit to RW~Tri. The main difference is the extension of the 
low-ionization region near the disk which is much smaller in this model. This 
is a consequence of the reduction in $\beta$ that was required to reproduce the
\ion{Si}{4} the absorption dips.

\begin{figure}[t]
  \centering
  \ifthenelse{\boolean{OnlineEdition}}{
    \includegraphics[width=0.48\textwidth]{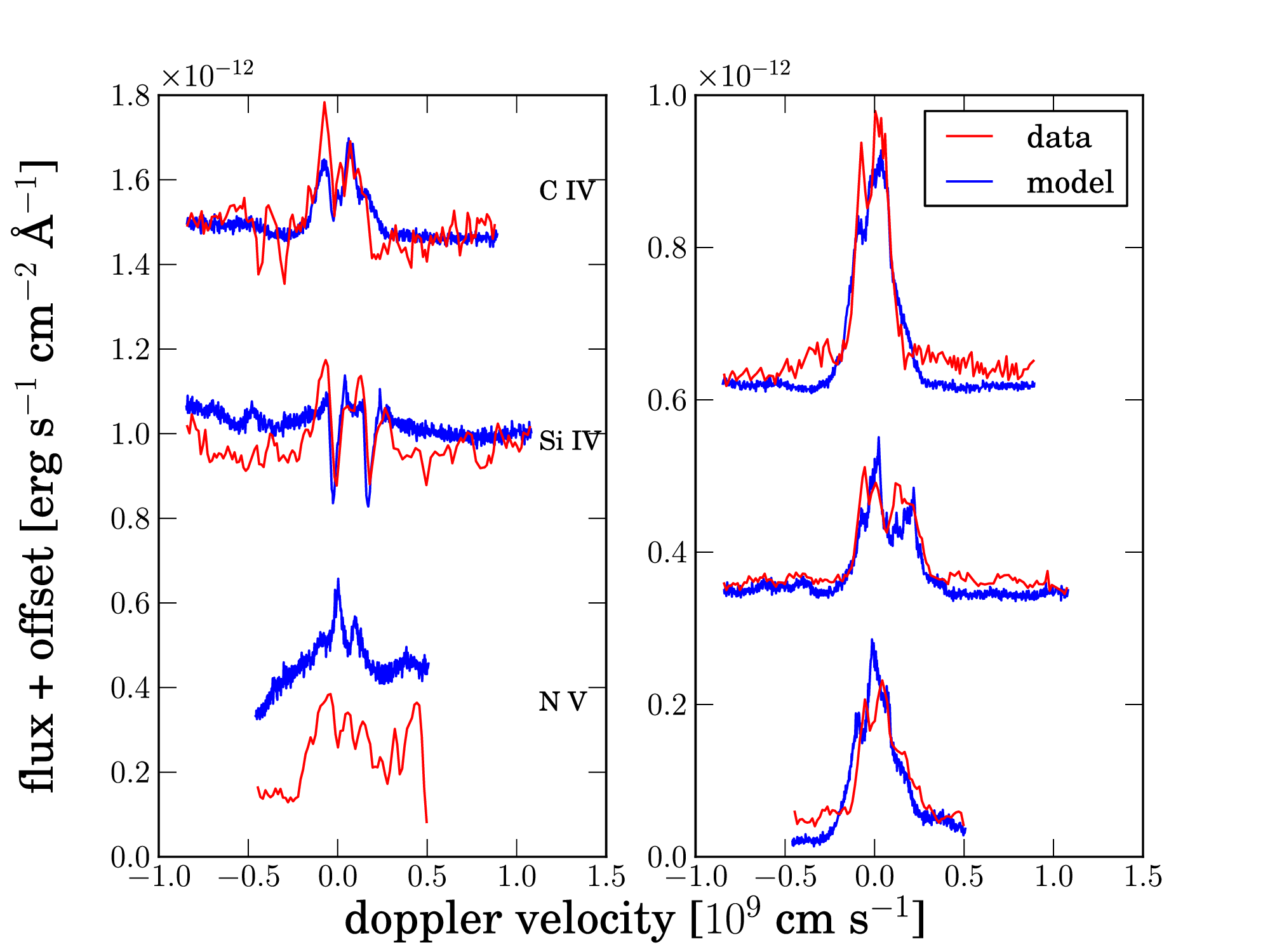}
  }
  {
    \includegraphics[width=0.48\textwidth]{uxuma_sv93_lines_1_bw.eps}
  }
  \caption{Details of line profiles for the UX UMa SV93 model during pre-eclipse (left) and mid-eclipse (right).}
  \label{fig:uxuma_sv_profiles}
\end{figure}

As for RW~Tri, we were able to transform the wind parameters to the
SV93 prescription and obtain a wind model which produced comparably
good spectra to the best KWD95 model (Figure \ref{fig:uxuma_sv_profiles}). 
Some minor differences appear in the profile shapes, in particular the 
\ion{Si}{4} absorption dips are slightly deeper. However, once again, the 
characteristic wind structure is qualitatively similar, as are the line-forming
regions.

\subsection{Similarities and Differences between the RW Tri and UX UMa wind Models}
\label{sec:Comparison}

Compared to previous studies in which the ionization state of the wind
has been imposed, our approach has the important advantage that the
ionization conditions in the wind are treated self-consistently with
the radiation field. This has directly led us to wind models with simple,
physically plausible ion distributions that are
characteristically similar for both the objects we have studied with
both wind prescriptions. The favored picture always involves a polar
concentration of \ion{N}{5}, equatorial \ion{Si}{4} (and lower ions),
and \ion{C}{4} filling most of the space between. This basic
configuration was able to reproduce the observations, despite the
differences in the UX UMa and RW Tri observed spectra. As one would
expect, the major line formation occurs around the regions where the
relevant ions are most populated and is concentrated within the
innermost few tens of WD radii.

Our studies of the two systems have favored mass loss rates which are
a few percent of the accretion rates (6\% and 8\%). Our models suggest
that these values are fairly robust --- simple changes in the mass-loss
rate by a factor of 2 (in either direction) leads to results which are strongly inconsistent with the observations.

The main differences between our models for RW Tri and UX UMa lie in
the details of the wind structure and are connected to the differences
in the KWD95 parameters $\alpha$ (mass loss exponent) and $\beta$
(acceleration exponent). In RW Tri, a small value of $\alpha$ was necessary to avoid under-producing \ion{N}{5}. 
The $\alpha$-parameter is less strongly constrained for our UX UMa point model
-- $\alpha=0.5$ agrees very well with the observations and
leads to a simple wind structure but higher or lower values cannot be
strongly excluded. Thus, our models do not rule out the possibility
that both systems have somewhat similar radial dependence of the
mass-loss rate per unit area.

\begin{figure}[t]
  \centering
  \includegraphics[width=0.48\textwidth]{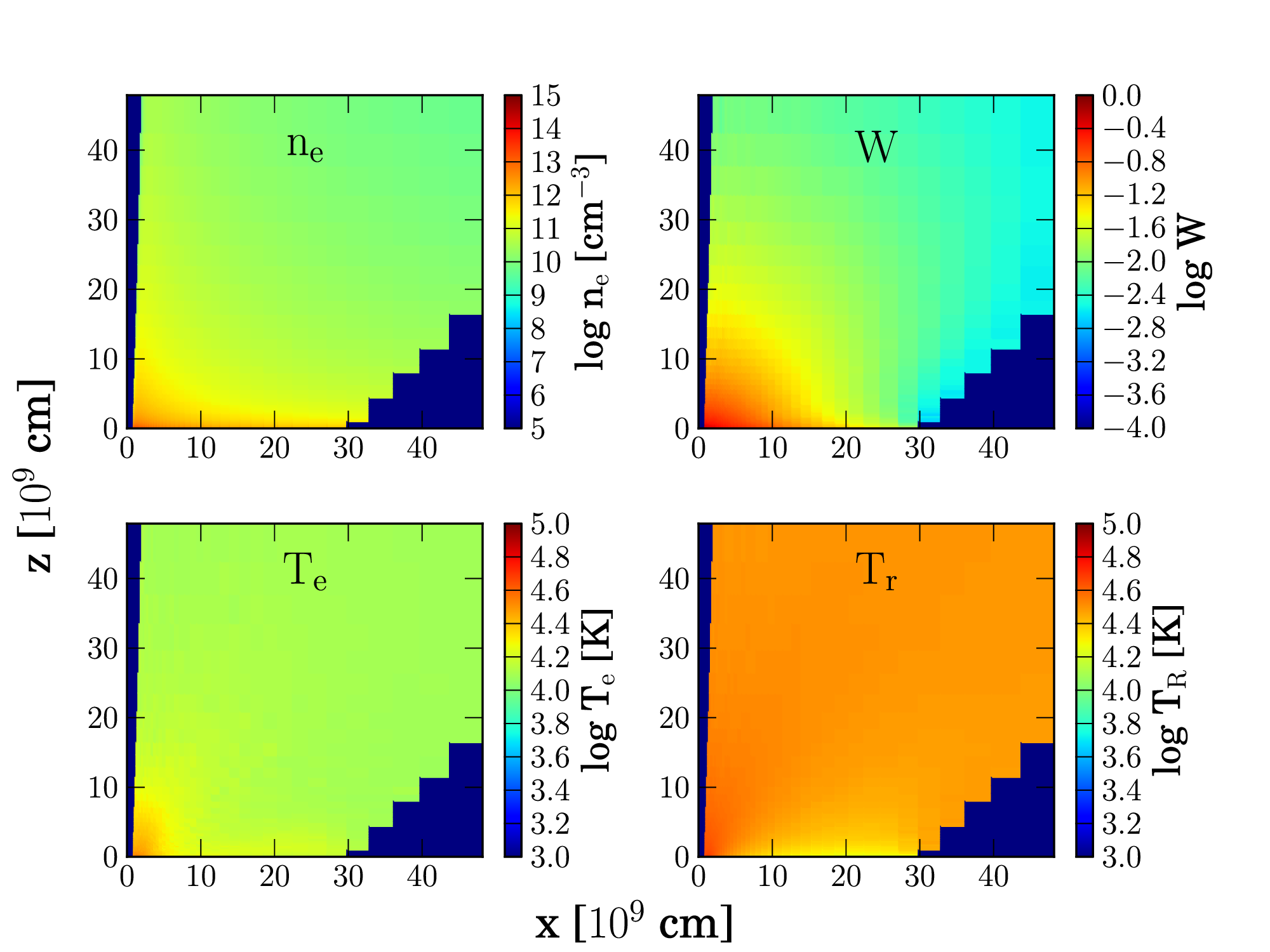}
  \caption{Temperature and density structure of the UX UMa KWD95 model.}
  \label{fig:uxuma_kwd_temp}
\end{figure}

On the other hand, our results relating to the wind acceleration law
do seem to suggest real differences between the systems. The $\beta$-parameter
cannot be very significantly reduced for RW~Tri because this would
eventually introduce strong absorption features in \ion{Si}{4}
and \ion{C}{4}, neither of which is observed. (The high acceleration
exponent ensures that the main \ion{Si}{4} region is located
far above the disk, outside the line of sight.) A large value of $\beta$ also
allows for a low-ionization state region close to the disk (below the
main \ion{Si}{4} concentration) where absorption lines seen
in the pre-eclipse spectra (e.g., \ion{C}{2}) can form. 
However, the UX~UMa spectra also place relatively
strong constraints on the wind velocity law for that system. Here, \ion{Si}{4}
absorption dips are seen, favoring a higher
ionization state closer to the disk compared to RW Tri. In our model,
this increase in the
ionization state was realized via a rapid acceleration (low
acceleration exponent), leading to a low density. 
We note that this is in contrast to the study of UX~UMa
by \cite{Knigge1997} in which a {\it high} $\beta$-value (i.e., {\it 
slow} acceleration) was invoked to account for absorption dips
in \ion{C}{4}. In their approach, a constant ionization fraction was
assumed meaning that deeper absorption arises from higher
densities. In our case, we have to account for the fact that the
density affects not only the optical depth directly but also the
ionization balance. The net effect of this is that {\it lower} densities
are needed to (see Figures \ref{fig:rwtri_kwd_temp} 
and \ref{fig:uxuma_kwd_temp})
obtain the \ion{Si}{4} dips because high densities lead to too low an
ionization state. Thus, we find that significantly faster acceleration
is required to model the UX~UMa spectra than those of RW~Tri, likely
pointing to a real physical difference in the manner in which the
outflows of the two systems are driven.

A downside of the reduced density around the base of the flow in our
UX~UMa model is that the low-ionization state material is
pushed very close to the disk. Thus, we lose any trace of emission
or absorption signatures from very low-ionization state material in
the model spectra. This suggests that a more complex wind structure
(e.g., clumping) may be
needed at the highest inclination to allow for the simultaneous
appearance of absorption features in
both relatively high ionization state lines (e.g., \ion{Si}{4}) and
significantly less ionized material (e.g., \ion{C}{2}). We generally
found that these classes of absorption features could not be
simultaneously created with the kinematic wind models we have considered.

An additional, less significant difference between the point models of
UX UMa and RW Tri is the degree of wind collimation. As noted above, there is a
certain freedom for this parameter in RW Tri since the spectra are not
very strongly affected by it. We can, however, exclude very low
collimation since it produces P-Cygni \ion{C}{4} profiles which are in
conflict with observation. In UX UMa, winds with very low collimation
can also be excluded for the same reason. But in this system, the
absorption features in \ion{Si}{4} give additional constraints on the
collimation. A moderately high wind collimation gives the correct
blueshift and width of these features, introduced by the slowly
accelerating wind.

\section{Conclusions}
\label{sec:conclusions}

We have presented point models for the winds in the eclipsing CVs
RW~Tri and UX~UMa which are able to reproduce the main characteristics
of the major UV resonance lines of \ion{N}{5}, \ion{Si}{4},
and \ion{C}{4} both in and out of eclipse. The ionization conditions
in our models are primarily determined by the temperature distribution
in the wind which, in turn, is determined by the effective temperature
distribution of the disk. Accounting for this, we found that agreement
with observation could be obtained for winds with a characteristic
structure in which \ion{N}{5} forms around the inner edge of the wind
(exposed to the hottest disk radiation), while \ion{Si}{4} is
restricted to a region around the base of the flow that extends along
the equatorial edge of the flow. The equatorial location
of \ion{Si}{4} allows for narrow absorption dips to appear in the line
profile, as observed in UX~UMa out of eclipse. The
formation of the resonance lines of both \ion{N}{5} and \ion{Si}{4} is
restricted to the innermost boundaries of the wind (within a few tens
of WD radii of the central object). The outer wind has ionization
state intermediate between these edge zones and it is here that 
the strong \ion{C}{4} line is formed. The wind volume reasonable for
this line is significantly more extended than for either of the other
major resonance lines but still concentrated within $\sim 50$ WD radii
of the central object. 

Although our focus has been the major resonance lines, our models
include some wind regions of even lower ionization state material
which are able to introduce some absorption features in the RW~Tri
pre-eclipse spectra due to, e.g., \ion{C}{2} and \ion{Si}{3}. In
agreement with observations, our models also predict weak \ion{C}{3}
emission during eclipse and some recombination emission from
both \ion{H}{1} and \ion{He}{2}. The details of these features are
less well reproduced than those of the main resonance lines: in
particular, \ion{H}{1} Ly$\alpha$ is significantly overpredicted
by our models for RW~Tri while \ion{He}{2} is dramatically underpredicted for
UX~UMa. The failures are most likely associated with the manner in which our 
wind models describe the very base of the wind where it is densest and, 
perhaps, has the most complex flow geometry. 

We have used two different classes of simply parameterized flow models (those 
due to KWD95 and SV93) and confirmed that both parameterization are able to 
reproduce the observations with comparable accuracy and that they lead to 
similar conclusions relating to the ionization and structure of the flow.

The wind structure our models suggest for both RW~Tri and UX~UMa are
characteristically similar. Both require wind mass-loss rates which
are $\sim$6\% -- 8\% of the accretion rate. We also found some
evidence suggesting that obtaining a
wind able to
simultaneously account for all three of the major resonance lines
may favor models in which the mass-loss per unit area of the disk is not
strongly coupled to the local disk effective temperature. In
particular, for RW~Tri we found that adopting $\alpha = 1$ for the
KWD95 prescription (meaning that the mass loss is proportional to the
local disk luminosity) significantly underpredicts the \ion{N}{5}
emission. To reproduce the observed \ion{N}{5} line strength, we found
that a much lower value of $\alpha \sim 0.1$ was
required. Qualitatively similar results were obtained from the SV93
prescription. In UX~UMa, we found that the relative line strengths were
less constraining on the distribution of mass loading in the wind but
are also consistent with $\alpha$-values somewhat less than 1.0. 
These conclusions are likely to be affected if there are any additional
sources of ionizing radiation (e.g., a hot boundary layer) that the
calculations presented here do not include. Nevertheless,
they may implicate mechanisms other than the radiation force alone in
launching the wind since it suggests the connection between the local
mass loss and luminosity in the disk may be weak. 

We found some evidence for real differences in the wind properties
between the two systems. In particular, our models suggest that the
acceleration around the base of the flow is significantly more rapid in
UX~UMa than in RW~Tri. This conclusion mainly stems from the presence
of absorption dips in the out-of-eclipse \ion{Si}{4} profile of UX~UMa
which are absent from RW~Tri. Our attempts to model these features as
part of the wind argue against a universal
kinematic outflow model and may suggest that there are real
differences in the wind launching regions of different
systems. However, further study of the structure around the base of the wind
(including consideration of lower ionization state spectral features)
is required to confirm this and we cannot yet rule out that the absorption dips
in \ion{Si}{4} might have some different origin (e.g., in the rim of
the accretion disk).

Overall, our results support conclusions drawn from
previous theoretically motivated considerations of the structure of CV
disk winds and their physical origin. In
particular, as discussed by \cite{Drew2000},
the mass-loss rates predicted from simulations of radiatively
driven outflows \citep{Proga1998} are likely too 
small to account for the strength of observed spectral features.
\citep{Drew2000,Proga2002}. Our results strongly
support this conclusion: our wind mass-loss rates are indeed significantly
higher than those expected for line-driven disk
wind models with a plausible value of the Eddington ratio (see
Figure 1 of \citealt{Drew2000}), while the disk accretion rates we have
used are close to (or
below) the lower limit of the regime in which radiation pressure alone
might be able to drive a flow. In addition, we have concluded that the
range of radii from which mass loss is launched is significantly more
extended than one might expect for a radiative driven flow (i.e., the
low $\alpha$-value which we require for RW~Tri). This is in line with
the discussion
of \cite{Proga2002} in which they also argue that line-driven disk wind
simulations do not succeed in launching fast outflows from as
extended a region of the disk as suggested by
observations. Thus, although the accretion
and mass-loss rates implied
by our models for the eclipsing nova-like variables are undoubtedly in
the range where radiation forces should be dynamically
important, additional physics --- most plausibly
magnetic phenomena --- likely has a part in driving the mass loss.

\acknowledgements{This work was supported by NASA through grant HST-AR-10674.01-A from the Space
Telescope Science Institute, which is operated by AURA, Inc., under
NASA contract NAS5-26555.  Most of the work presented here was conducted during U.M.N.'s
research visit at STScI. U.M.N.\ thanks K.S.L.\ for his
encouraging support and kind hospitality during that time. S.A.S.\
acknowledges STScI support for a collaborative visit during which part
of this project was performed.}

\newpage






\newpage


\end{document}